\documentstyle [10pt,eqsecnum,aps] {revtex}
\input epsf
\newcommand{\beq}{\begin{equation}}
\newcommand{\eeq}{\end{equation}}
\newcommand{\beqs}{\begin{eqnarray}}
\newcommand{\eeqs}{\end{eqnarray}}

\newcommand{\Z}{\mbox{\rm Z$\!\!$Z}}

\begin{document}

\draft

\baselineskip 6.0mm

\title{Chromatic Polynomials for Families of Strip Graphs and their Asymptotic 
Limits} 

\vspace{4mm}

\author{Martin Ro\v{c}ek\thanks{email: rocek@insti.physics.sunysb.edu} 
\and Robert Shrock\thanks{corresponding author, 
email: shrock@insti.physics.sunysb.edu}
\and Shan-Ho Tsai\thanks{email: tsai@insti.physics.sunysb.edu}}

\address{
Institute for Theoretical Physics  \\
State University of New York       \\
Stony Brook, N. Y. 11794-3840}

\maketitle

\vspace{4mm}

\begin{abstract}

    We calculate the chromatic polynomials $P((G_s)_m,q)$ and, from these, the
asymptotic limiting functions $W(\{G_s\},q)=\lim_{n \to \infty}P(G_s,q)^{1/n}$ 
for families of $n$-vertex graphs $(G_s)_m$ comprised of $m$ repeated subgraphs
$H$ adjoined to an initial graph $I$.  These calculations of $W(\{G_s\},q)$ for
infinitely long strips of varying widths yield important insights into
properties of $W(\Lambda,q)$ for two-dimensional lattices $\Lambda$.  In turn,
these results connect with statistical mechanics, since $W(\Lambda,q)$ is the
ground state degeneracy of the $q$-state Potts model on the lattice $\Lambda$. 
For our calculations, we develop and use a generating function method, which
enables us to determine both the chromatic polynomials of finite strip graphs
and the resultant $W(\{G_s\},q)$ function in the limit $n \to \infty$. From
this, we obtain the exact continuous locus of points ${\cal B}$ where 
$W(\{G_s\},q)$ is nonanalytic in the complex $q$ plane.  This locus is shown
to consist of arcs which do not separate the $q$ plane into disconnected 
regions.  Zeros of chromatic polynomials are computed for finite strips and 
compared with the exact locus of singularities ${\cal B}$.  We find that as 
the width of the infinitely long strips is increased, the arcs comprising 
${\cal B}$ elongate and move toward each other, which enables one to 
understand the origin of closed regions that result for the (infinite) 2D 
lattice. 

\end{abstract}


\vspace{10mm}

\pagestyle{empty}
\newpage

\pagestyle{plain}
\pagenumbering{arabic}
\renewcommand{\thefootnote}{\arabic{footnote}}
\setcounter{footnote}{0}

\section{Introduction}

    Nonzero ground state disorder and associated entropy, $S_0 \ne 0$, is an
important subject in statistical mechanics.  One physical example is provided
by ice, for which the residual molar entropy is
$S_0 = 0.82 \pm 0.05$ cal/(K-mole), i.e., $S_0/R = 0.41 \pm 0.03$, where
$R=N_{Avog.}k_B$ \cite{ice,lp,liebwu,ps,atkins}.  
Indeed, residual entropy at low
temperatures has been observed in a number of substances, including nitrous
oxide, NO, carbon monoxide, CO, and FClO$_3$ (a comprehensive review is given
in Ref. \cite{ps}).  In these examples, the entropy occurs without 
frustration.  In magnetic systems, two examples are provided by the
Ising antiferromagnet on the triangular and kagom\'e lattices
\cite{wannier,kn}; here, the ground state entropy does involve frustration.
One should also mention the recent experimental observation of compounds 
whose behavior can be modelled by quantum Heisenberg antiferromagnets on a
kagom\'e lattice, 
including SrCr$_{8-x}$Ga$_{4+x}$O$_{19}$, (SCGO(x)) \cite{scgo} and 
deuteronium jarosite, (D$_3$O)Fe$_3$(SO$_4)_2$(OD)$_6$ \cite{dj}.  The 
quantum Heisenberg antiferromagnet has a disordered ground state
with finite entropy and frustration on the kagom\'e lattice \cite{hafkag}. 
A particularly simple model exhibiting
ground state entropy without the complication of frustration is the $q$-state
Potts antiferromagnet (AF) \cite{potts,wurev} on a lattice $\Lambda$, for
$q \ge \chi(\Lambda)$, where $\chi(\Lambda)$ denotes the minimum number of
colors necessary to color the vertices of the lattice such that no two adjacent
vertices have the same color.  As is already evident from the foregoing, this
model also has a deep connection with graph theory since the zero-temperature 
partition function of the above-mentioned $q$-state Potts antiferromagnet on 
a lattice $\Lambda$ satisfies $Z(\Lambda,q,T=0)_{PAF}=P(\Lambda,q)$, where 
$P(G,q)$ is the chromatic
polynomial expressing the number of ways of coloring the vertices of a graph
$G$ \footnote{We recall the strict mathematical definition of a graph as 
a collection of vertices and bonds connecting (all or a subset of the)
vertices, such that there are (i) no loops, {\it i.e.}, bonds connecting a 
vertex to
itself, (ii) no multiple bonds connecting a given pair of vertices, and (iii)
the total number of vertices is finite.  Thus, an infinite lattice is not a
graph, but a limit thereof. A graph is termed empty if it has no bonds
connecting the vertices.}
with $q$ colors such that no two adjacent vertices (connected by a bond of
the graph) have the same color \cite{birk,whit,bl,rtrev,tutterev,biggsbook}.  
Hence, the ground state entropy per site is
given by $S_0/k_B = \ln W(\Lambda,q)$, where $W(\Lambda,q)$, the ground state
degeneracy per site, is
\beq
W(\Lambda,q) = \lim_{n \to \infty} P(\Lambda_n,q)^{1/n}
\label{w}
\eeq
Here, $\Lambda_n$ denotes an $n$-vertex lattice of type $\Lambda$ (square,
triangular, etc.), with appropriate (e.g., free) boundary conditions.

    Since nontrivial exact solutions for this
function are known in only a very few cases (square lattice for $q=3$
\cite{lieb}, triangular lattice \cite{baxter}, and kagom\'e lattice for $q=3$
\cite{wurev}), it is important to exploit and extend general approximate
methods that can be applied to all cases.
Such methods include rigorous upper and lower
bounds, large-$q$ series expansions, and Monte Carlo measurements.  Recently,
we studied the ground state entropy in antiferromagnetic Potts models on
various lattices and obtained further results with these three methods
\cite{w,w3,wa,wn}; see also Ref. \cite{sokal}.  As $P(G,q)$ is a 
polynomial, $q$ may be continued from integer to complex values, which we do 
here. 

   One can gain valuable insight into the values of the ground state degeneracy
of the Potts antiferromagnet $W(\Lambda,q)$ for different 2D lattices 
$\Lambda$ and values of $q$ by studying this function on infinitely long strips
of increasing widths.  Accordingly, in this paper, we calculate 
$P((G_s)_m,q)$ for a family of strip 
graphs $(G_s)_m$ which are comprised of $m$ repetitions of a subgraph unit 
$H$ attached to an initial subgraph $I$ on one end (which, by convention, we 
take to be the right end).  Symbolically, we denote such a strip graph as 
\beq
(G_s)_m = (\prod_{\ell=1}^{m} H)I
\label{gs}
\eeq
for $m \ge 1$, with $(G_s)_0 \equiv I$.  The initial subgraph $I$ 
may itself be equal to $H$.  To carry out these calculations, we construct 
a generating function $\Gamma(G_s,q,x)$ that yields the 
chromatic polynomial $P((G_s)_m,q)$ as the coefficient of the term $x^m$ in 
its Taylor series expansion about $x=0$, where $x$ is a formal auxiliary
variable.  We then use this generating function to compute 
the asymptotic limiting function for the infinitely long strip graph of type 
$G_s$, $W(\{G_s\},q)$ defined in eq.\ (\ref{w}), and to characterize its 
analytic properties in the complex $q$ plane.  We do not connect the ends of 
the strip in the longitudinal direction, {\it i.e.}, the strip has free 
boundary conditions in both the transverse and longitudinal
directions. As in Refs. \cite{w,w3,wa,wn}, we let ${\cal B}$ denote the 
continuous locus of points in the $q$ plane where $W(\{G_s\},q)$ is 
nonanalytic.  
For the infinite-length limit of the strip graphs $G_s$ defined in eq.\ 
(\ref{gs}) this locus of points ${\cal B}$ consists of arcs which do not 
separate the complex $q$ plane into different regions.  We calculate these 
for a number of different types of strips.  

In general, in the limit as the number $n$ of vertices on an $n$-vertex 
graph of type $G$ 
goes to infinity, the locus ${\cal B}$ forms as a merging together of zeros 
of the chromatic polynomial $P(G,q)$ (called chromatic zeros of $G$).  
It is thus of interest to calculate chromatic zeros 
for strip graphs $(G_s)_m$ of finite length and to compare these with the 
asymptotic limit represented by the curves making up ${\cal B}$.  We perform 
this comparison for a number of types of strip graphs here.  

   Our results are useful for several reasons.  First, they are a set of 
exact calculations of chromatic polynomials for a large variety of strips, of
both finite and infinite length.  Second, the generating function 
introduced here enables us to calculate, in an 
algorithmic manner, the chromatic polynomials for very long finite strips. 
To appreciate how valuable such a method is, we recall the well-known fact 
\cite{rtrev} that, in general, 
the calculation of the chromatic polynomial of an arbitrary graph is an 
NP-complete problem (NP = nonpolynomial), {\it i.e.}, 
roughly speaking, it requires 
a time which grows exponentially as a function of the number of vertices in 
the graph.  Third, the generating function provides an elegant way to obtain 
directly the infinite-length limiting function $W(\{G_s\},q)$ for a given strip
graph of type $G_s$.  These $W(\{G\},q)$ 
functions show a rich and fascinating analytic structure in the complex $q$
plane.  Fourth, as noted before, 
by studying this structure as a function of the width of the
strips, one gains insight into how the limit of the infinite 2D lattice is
approached (with free boundary conditions).  

   We define 
\beq
W(\{G\},q) \equiv W(\{G\},q_s)_{D_{qn}} \equiv 
\lim_{q \to q_s} \lim_{n \to \infty} P(G,q)^{1/n}
\label{wdefqn}
\eeq
and recall from Ref. \cite{w} that at certain special points $q_s$, the 
following limits do not commute: 
\beq
\lim_{n \to \infty} \lim_{q \to q_s} P(G,q)^{1/n} \ne
\lim_{q \to q_s} \lim_{n \to \infty} P(G,q)^{1/n}
\label{wnoncomm}
\eeq
Our present definition (\ref{wdefqn}) is the same as in Ref. \cite{w} and has
the advantage of maintaining the analyticity of $W(\{G\},q)$ at the special
points $q_s$.  

   In Ref. \cite{w} two of us had also discussed a second 
subtlety in the definition of
$W(\{G\},q)$: for certain ranges of real $q$, $P(G,q)$ can be negative, and, 
of course, when $q$ is complex, so is $P(G,q)$ in general. In these cases it 
may not be obvious, {\it a priori}, which of the $n$ roots
\beq
P(G,q)^{1/n} = \{ |P(G,q)|^{1/n}e^{2\pi i r/n} \} \ , \quad r=0,1,...,n-1
\label{pphase}
\eeq
to choose in eq.\ (\ref{w}). 
Consider the function $W(\{G\},q)$ defined via eq.\ (\ref{w})
starting with $q$ on the positive real axis where $P(G,q) > 0$, and consider
the maximal region in the complex $q$ plane that can be reached by analytic
continuation of this function.  We denote this region as $R_1$.  Clearly, the
phase choice in (\ref{pphase}) for $q \in R_1$ is that given by $r=0$, namely
$P(G,q)^{1/n} = |P(G,q)|^{1/n}$.  For the various exactly solved cases in 
Ref. \cite{w} concerned with families of graphs $\{G\}$ for
which the areas of analyticity of $W(\{G\},q)$ include other regions not
analytically connected to $R_1$, there is not, in general, any canonical 
choice of phase in (\ref{pphase}).  However, for the strip graphs studied 
here, the curves ${\cal B}$ where $W(\{G_s\},q)$ is nonanalytic do not 
separate the $q$ plane into different regions, {\it i.e.}, the entire $q$ 
plane, except for the locus ${\cal B}$, is in
$R_1$, so this second point having to do with phase choice in (\ref{pphase}) 
is easily dealt with, as discussed below. 

   This paper is organized as follows.  In Section II we introduce the 
generating function.  In Section III we calculate the
generating functions for a number of interesting types of strip graphs $G_s$.
Section IV is devoted to a study of the analytic properties of the resultant
$W(\{G_s\},q)$ functions.  Some final remarks are given in the conclusions. 
Besides the mathematical references 
\cite{birk,whit,bl,rtrev,tutterev,biggsbook} and the previous 
Refs. \cite{w,w3,wa,wn} by two of us, some related work is in Refs. 
\cite{lieb,bds,bkw,baxter,read91}.   We shall report further results in 
Ref. \cite{strip2}. 

\section{Calculation of $P(G_{\lowercase{s}},\lowercase{q})$ via a 
Generating Function}

\subsection{Generalities}

   It is possible to calculate the chromatic polynomials for
a wide variety of strip graphs $(G_s)_m$ of the type defined in eq.\ (\ref{gs})
by computing a generating function.  
This also enables us immediately to calculate the resultant asymptotic 
limiting function $W(\{G_s\},q)$.  Before proceeding, we discuss some
preliminaries.  We denote the number of vertices in a graph $G$ as $n(G)$. 
Consider two successive repeating subgraph units $H$ and denote
their intersection as the graph
\beq
L_H = H_{r+1} \cap H_{r}
\label{l}
\eeq
Let the number of vertices in the repeating subgraph unit $H$, the intersection
graph $L_H$, and the initial graph $I$ be $n(H)$, $n(L_H)$, and $n(I)$,
respectively.  Then the total number of vertices $n$ in the strip graph 
$(G_s)_m$ is 
\beq
n((G_s)_m)=c_{s,1}m+c_{s,0}
\label{nvertices}
\eeq
where 
\beq
c_{s,0}=n(I)
\label{cs0}
\eeq
and 
\beq
c_{s,1} = n(H)-n(L_H)
\label{cs1}
\eeq
The values of these coefficients are listed in Tables I and II for the various 
strip graphs that we consider. 

\begin{figure}
\centering
\leavevmode
\epsfxsize=4.0in
\epsffile{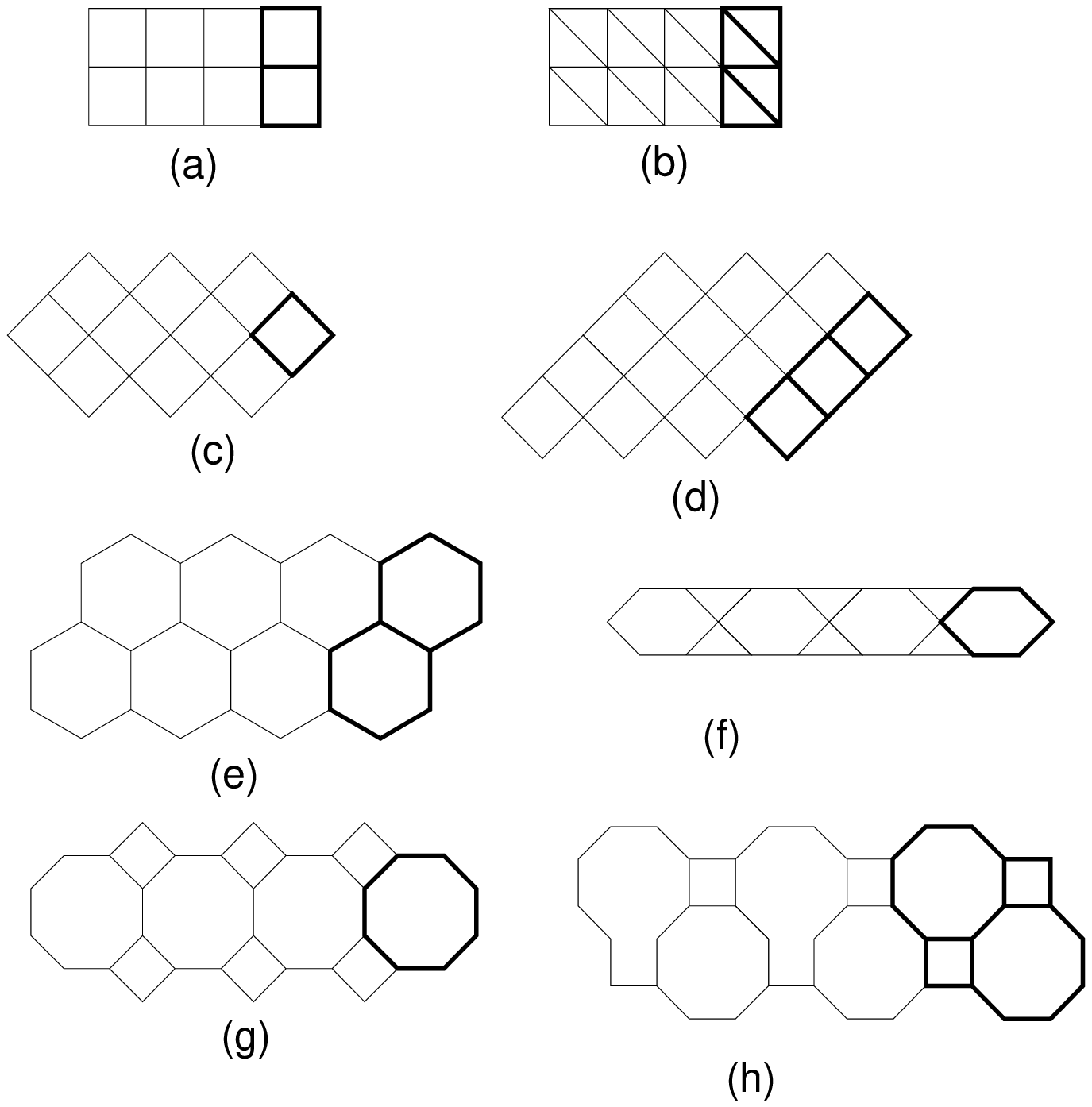}
\vspace{-2cm}
\caption{\footnotesize{
Illustrations of strip graphs $(G_{s(L_y)})_m$ of type $s$. Subscripts
$L_y$ and $m$ refer, respectively, to the width and the number of repeated 
subgraph units $H$ beyond the initial subgraph $I$ (where $I$ may be identical
to $H$).  For (a)-(g), $m=3$, whereas $m=2$ for (h). For ease of 
identification, the initial subgraph $I$ is drawn with heavier lines.  
See text for further discussion of each type of strip graph. 
(a) $(G_{sq(3)})_3$, an $L_x \times L_y = 5 \times 3$ strip of the square 
lattice; 
(b) $(G_{tri(3)})_3$, an $L_x \times L_y = 5 \times 3$ strip of the 
triangular lattice (represented as a square lattice with additional oblique 
bonds); 
(c) $(G_{sq_d(3)})_3$, a strip of a square lattice along the diagonal
direction, with the repeating subgraph $H$ comprised of the upper, lower, and 
left members of a $3 \times 3$ diagonally oriented box; 
(d) $(G_{sq_d(4)})_3$, as in (c), but with a repeating unit $H$ comprised of 
a diagonally oriented strip of 4 squares; 
(e) $(G_{hc(3)})_3$, a strip of the honeycomb lattice; 
(f) $(G_{kag(2)})_3$, a strip of the kagom\'e lattice; 
(g) $(G_{488(2)})_3$, a strip of the $(4 \cdot 8^2)$ lattice; 
(h) $(G_{488(3)})_2$, a wider strip of the $(4 \cdot 8^2)$ lattice along a
direction rotated $45^\circ$ relative to that in (g).}}

\label{figstrip}
\end{figure}

For example, consider strips of the square, triangular, and honeycomb 
lattices, as shown in Figs.\ \ref{figstrip}(a,b,e). We represent the strip of 
the triangular lattice as a strip of squares with oblique bonds added, as 
indicated in Fig.\ \ref{figstrip}(b).  For our present discussion we 
represent the honeycomb lattice of Fig.\ \ref{figstrip}(e) as a brick lattice. 
Denote the length as $L_x$ vertices in the horizontal
(longitudinal) direction, and the width as $L_y$ vertices in the vertical 
(transverse) direction.  This corresponds to a width of $w = L_y-1$ layers 
of squares or hexagons and $2w$ triangles, for the respective lattices. 
The strips of the square and triangular lattices shown in Fig.\ 
\ref{figstrip}(a,b) have $L_x=5$ and $L_y=3$. 
These strips are thus $L_x \times L_y$ lattices 
with free boundary conditions in both $x$ and $y$ directions. 
The strip of the honeycomb lattice of Fig.\ \ref{figstrip}(e), when represented
in brick form, has $L_x=9$ for the top and bottom edges 
and $L_x=10$ for the interior layer, with $L_y=3$. 
Later we will study the limit in which $L_x \to \infty$ with $L_y$ fixed,
{\it i.e.}, strips of size $\infty \times L_y$.  
The square strip has, as a repeating subgraph unit $H$, a vertical 
stack of $w$ squares.  Hence, starting at one end, which will be taken by 
convention to be on the right and to consist of an initial subgraph $I$ which
is identical to $H$, the elongation of the strip proceeds by the
addition of successive vertical stacks on the growing (left) end. For the
honeycomb and triangular lattices, respectively, $H$ is an oblique stack of $w$
hexagons and a vertical stack of $2w$ triangles, and the initial subgraph
$I=H$.  For the strips of the square lattice (along the row direction) and of
the triangular lattice, the relation between $L_x$ and $m$ is thus 
$L_x=m+2$, and 
\beq
n(H) = 2n(L_H) = 2L_y  \ \quad {\rm for} \quad G_{sq(L_y)} \quad {\rm or} 
\quad G_{t(L_y)}
\label{nhsqt}
\eeq
while for the honeycomb lattice, 
\beq
n(H) = 2(2L_y-1) \ , \qquad n(L_H) = 2(L_y-1) \qquad {\rm for} \quad 
G_{hc(L_y)}
\label{nhhc}
\eeq
The number $n(I)$ depends on one's convention for the initial subgraph; in
these three types of strips, since we take $I=H$, we have $n(I) = n(H)$ for
each.  The total number of vertices is 
$n((G_{s(L_y)})_m)=L_xL_y=(m+2)L_y$ for the square and triangular lattices, 
so 
\beq
c_{s(L_y),1} = L_y \ , \ \quad c_{s(L_y),0} = 2L_y \ \quad {\rm for} 
\quad s=sq, \ t
\label{cs10sqt}
\eeq
while for the honeycomb (brick) lattice, $n((G_{hc(L_y)})_m)=2[(m+2)L_y-1]$, so
\beq
c_{hc(L_y),1} = 2L_y \ , \ \quad c_{hc(L_y),0} = 2(2L_y-1)
\label{cs10hc}
\eeq
It is straightforward to obtain the analogous formulas for other types of strip
graphs $G_s$; we list the results for $c_{s,1}$ and $c_{s,0}$ in Tables I and
II. 

\subsection{Main Theorem and Proof}

The generating function for a strip of type
$G_s$, denoted generically as $\Gamma(G_s,q,x)$, yields the chromatic 
polynomials via the expansion 
\beq
\Gamma(G_s,q,x) = \sum_{m=0}^{\infty} P((G_s)_m,q)x^m
\label{gamma}
\eeq
where $x$ is a symbolic variable and $P((G_s)_m,q)$ is the chromatic 
polynomial for the coloring, with $q$ colors, of the strip $(G_s)_m$ with 
$m$ repeated subgraph units $H$ beyond the initial ($m=0$) one, $I$. 
We find that the generating function $\Gamma(G_s,q,x)$ is a rational function 
of the form 
\beq
\Gamma(G_s,q,x) = \frac{{\cal N}(G_s,q,x)}{{\cal D}(G_s,q,x)}
\label{gammagen}
\eeq
with
\beq
{\cal N}(G_s,q,x) = \sum_{j=0}^{j_{max}} a_{s,j}(q) x^j
\label{n}
\eeq
and
\beq
{\cal D}(G_s,q,x) = 1 + \sum_{k=1}^{k_{max}} b_{s,k}(q) x^k 
\label{d}
\eeq
where the $a_{s,i}$ and $b_{s,i}$ are polynomials in $q$ that depend on the 
type and width of the strip graph $G_s$.  The polynomials $a_{s,i}$ also 
depend on the subgraph forming the right-hand end 
of the strip (discussed further below), while the $b_{s,i}$ are independent 
of this subgraph.  The respective degrees $j_{max}$ and $k_{max}$ of the 
numerator and denominator polynomials ${\cal N}(G_s,q,x)$ and 
${\cal D}(G_s,q,x)$, as functions of $x$, depend on the type and width of the 
strip $G_s$. 

   We can state these results formally: 

\vspace{4mm}

\begin{flushleft}
Theorem 1. 
\end{flushleft}

Let a strip graph of type $G_s$ be constructed by $m$ successive additions of 
a repeating graphical subunit $H$ to an initial subgraph $I$ as in
eq.\ (\ref{gs}). Then the chromatic polynomial for this strip graph can be 
expressed via the generating function (\ref{gamma}) given by
(\ref{gammagen})-(\ref{d}). 

\vspace{4mm}

\begin{flushleft}
Proof
\end{flushleft}

We use the deletion-contraction theorem, which states that the chromatic
polynomial for a graph $G$ containing two adjacent vertices $v$ and $v'$ 
connected by a bond (edge) is equal to the chromatic polynomial for 
a graph in which this bond is removed minus the chromatic polynomial for
the graph in which these vertices are identified\footnote{
The deletion-contraction theorem is an 
expression of the elementary fact the chromatic polynomial of $G$ 
enumerates the coloring of the graph such that the colors of each pair of 
adjacent vertices are different.  In particular, then, the number of colorings
in which the colors on the pair $v$ and $v'$ are different is equal to the
number of colorings in which the colors assigned to $v$ and $v'$ are allowed 
to vary freely (subject to the constraints from their other bonds) minus the 
number in which the colors assigned to $v$ and $v'$ are the same.  Reading this
equation in the opposite direction gives the addition-contraction theorem of
graph theory, which we shall use later.  See, e.g., 
Refs. \cite{rtrev}-\cite{biggsbook} for further discussion of the mathematical
background.}. 
Applying the deletion-contraction theorem to the initial subgraph $I$, 
we obtain a finite set of linear equations, with $m$-independent coefficients, 
for $P((G_s)_m,q)$, $P((G_s)_{m-1},q)$, $P((G_{s,e})_m,q)$, and possibly 
$P((G_{s,e})_{m-1},q)$, where $G_{s,e}$ denotes a strip graph of type $s$ but 
with a different end subgraph (e) at the starting point on the right. 
In general, the application of the deletion-contraction theorem yields 
several $G_{s,e}$'s with different end subgraphs $e$.  The set of equations is
linear because the deletion-contraction theorem is a linear equation among 
chromatic polynomials.  The set of linear equations is finite because the 
repeating subgraph unit $H$ is a finite graph.  Next, assuming the form 
(\ref{gamma}) for each type of strip and substituting into these equations, 
we obtain a resultant set of linear equations for the generating function of 
the original strip, $\Gamma(G_s,q,x)$, and the set of generating functions 
for the strips with
different ends produced by the action of the deletion-contraction theorem, 
$\Gamma(G_{s,e},q)$, which we solve.  This completes the proof.  \ $\Box$ 

\vspace{3mm}

   We illustrate the method of proof by an explicit example.  We
first note that a graph in which each adjacent pair of repeating subgraph
units $H$ intersect each other in a complete graph\footnote{
The complete graph $K_\ell$ is defined as the graph consisting 
of $\ell$ vertices, each of which is connected to all of the others with 
bonds.} $L_H = K_\ell$,
has a very simple chromatic polynomial.  Since the intersection of $H$ with
$L_H$ is just $L_H$, the intersection theorem\footnote{
The intersection theorem states that if two graphs $G$ and $G'$
intersect in a complete graph $K_\ell$ for some $\ell$, then the chromatic
polynomial of the union of these graphs is given by 
$P(G \cup G',q) = P(G,q)P(G',q)/P(K_\ell,q)$.} in graph theory
implies that the chromatic polynomial $P(H,q)$ factorizes according to 
\beq
P(H,q) = P(K_\ell,q)P_r(H,q)
\label{phkell}
\eeq
where
\beq
P(K_\ell,q)=\prod_{t=0}^{\ell-1}(q-t)
\label{pkell}
\eeq
which defines the polynomial $P_r(H,q)$.  For a chain graph 
$(G_{s;K_\ell})_m$ consisting of $m$ repeated subgraphs $H$ with an initial 
subgraph $I=H$, the chromatic polynomial is easily computed by the interative 
use of the intersection theorem:  
\beq
P((G_{s;K_\ell})_m,q) = \frac{P(H,q)^{m+1}}{P(K_\ell,q)^m} = 
\Bigl ( \prod_{t=0}^{\ell-1} (q-t) \Bigr ) P_r(H,q)^{m+1}
\label{pghfactor}
\eeq
The generating function is 
\beq
\Gamma(G_{s;K_\ell},q,x) =  \frac{P(H,q)}{1-P_r(H,q)x}
\label{gammakell}
\eeq
For example, for a strip of $m$ $p$-sided polygons ($p$-gons) with each 
adjacent pair of $p$-gons intersecting on a mutual edge ($K_2$), the 
chromatic polynomial is 
\beq
P((G_{p-gon;K_2})_m,q) = q(q-1)D_p(q)^m
\label{ppgon}
\eeq
where
\beq
D_p(q) = \frac{P(C_p,q)}{q(q-1)} =
\sum_{s=0}^{p-2}(-1)^s {{p-1}\choose {s}} q^{p-2-s}
\label{dk}
\eeq
and the chromatic polynomial for a circuit with $p$ vertices is
\beq
P(C_p,q) = (q-1)^p + (-1)^p(q-1)
\label{pck}
\eeq
so that $D_3(q)=q-2$ and $D_4(q)=q^2-3q+3$.  (In the following, for brevity, 
we shall generally write out the expression for $D_3(q)$ but not for $D_k(q)$ 
for higher values of $k$.) 
Since the chromatic polynomial can easily be calculated 
for this type of strip without the use of our more powerful methods, and since,
as will be discussed below, the points in the complex $q$ plane where the 
resultant $W(\{G_s\},q)$ function is nonanalytic are at most isolated branch
points, so that ${\cal B}=\emptyset$, {\it i.e.}, the null set, we concentrate
on more complicated types of strip graphs (e.g. of greater width) where the 
factorization in eq.\ (\ref{pghfactor}) does not occur.  

\begin{figure}
\centering
\leavevmode
\epsfxsize=4.0in
\epsffile{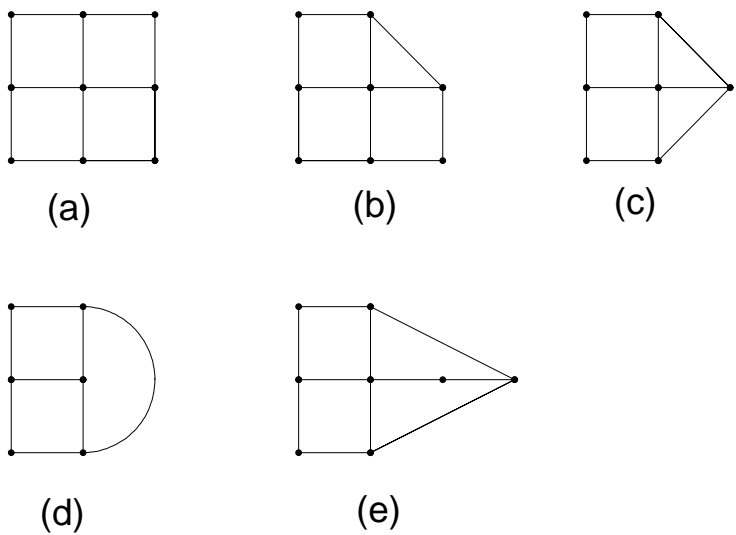}
\caption{\footnotesize{Graphs for square strips of width $L_y=3$ of type 
(a) $G_{sq,m=1}$ (b) $G_{sq,e=st,m=1}$, (c) $G_{sq,e=tt,m=1}$, 
(d) $G_{sq,e=t,m=1}$, and (e) $G_{sq,e=ttv,m=1}$.  See text for discussion.}}
\label{figsqexample}
\end{figure}

  Let us, then, illustrate our method of proof by considering a strip graph
that consists of squares, $G_{sq}$.  The method consists of using the
deletion-contraction theorem on the end of the original strip graph, which 
generates also related strip graphs with different initial subgraphs; one then
uses the deletion-contraction on these, thereby generating further types of
strip graphs, and so forth until no further additional types of strip graphs 
are 
generated.  (Of course, all of these additional types of strip graphs also have
free boundary conditions.)  In this manner, one obtains a set of linear 
equations for the generating functions of all of these types of strip graphs,
which is then solved.  A general strip of this type has size 
$L_x \times L_y$ vertices, {\it i.e.}, $(m+1) \times w$ squares, where 
$m=L_x-2$ and $w=L_y-1$, as given above.  Both the strip with $L_y=1$ (the
line) and $L_y=2$ have chromatic polynomials that factorize, so we consider 
the next simplest case of width $L_y=3$ (see Fig.\ \ref{figsqexample}(a)). 
 To simplify the notation as much as possible, we shall use the symbols 
$P_{s(L_y,e),m} \equiv P((G_{s(L_y),(e)})_m,q)$ and shall put 
$G_{sq(L_y=3)} \equiv G_{sq(3)} \equiv G_{ss}$.  
A first application of the deletion-contraction theorem on one of 
the vertical bonds (say, the upper one) at the right end of a length-$m$ 
strip yields the equation
\beq
P_{ss,m} = (q-1)D_4(q)P_{ss,m-1} - P_{st,m}
\label{sq1}
\eeq
for $m \ge 1$, where $P_{st,m}$ refers to the chromatic polynomial for the 
graph $G_{sq,e=st,m}$ with right-hand end consisting of a square and triangle
(whence the subscript $st$) shown in Fig.\ \ref{figsqexample}(b).  
Given our convention of taking
$G_{sq,m=0}$ to be the graph with two squares on top of each other and 
$G_{sq,e=st,m=0}$ to be a graph with a triangle adjoined to a square, it
follows that 
\beq
P_{ss,0} \equiv P((G_{sq(2)})_0,q) = q(q-1)D_4(q)^2
\label{pss0}
\eeq
and 
\beq
P_{st,0} = q(q-1)(q-2)D_4(q)
\label{pst0}
\eeq
Equation (\ref{sq1}) thus relates the chromatic polynomial of the square strip 
of length $m$ and width 2 to the chromatic polynomial of the same type of 
strip of length $m-1$ and to the chromatic polynomial of this type of square
strip with a different type of endgraph, as shown in Fig.\ 
\ref{figsqexample}(b).  We next apply the deletion-contraction theorem to 
this latter type of strip, specifically to the remaining vertical end bond 
on the right in Fig.\ \ref{figsqexample}(b), to obtain the
relation 
\beq
P_{st,m}=(q-1)(q-2)P_{ss,m-1} - P_{tt,m}
\label{sq2}
\eeq
for $m \ge 1$, and $P_{tt,m}$ is the chromatic polynomial for the square strip
graph $G_{sq,e=tt,m}$ with a right-hand end consisting of two triangles, as 
shown in Fig.\ \ref{figsqexample}(c).  
Here we take $G_{sq,e=tt,m=0}$ to be the two triangles
adjoined along their horizontal edge, so that 
\beq
P_{tt,0}=q(q-1)(q-2)^2
\label{ptt0}
\eeq
Proceeding by applying the theorem to one of the hypotenuses of the triangles 
in $G_{sq,e=tt,m}$, we derive
\beq
P_{tt,m} = (q-2)P_{ss,m-1} - P_{t,m}
\label{sq3}
\eeq
for $m \ge 1$, where the graph $G_{sq,e=t,m=1}$ is shown in
Fig.\ \ref{figsqexample} (d), so that
the $m=0$ graph of this type is a single triangle, with 
$P_{t,0}=P(K_3,q) = q(q-1)(q-2)$.  The next deletion, on the round bond on 
the right in Fig.\ \ref{figsqexample}(d), yields 
\beq
P_{t,m} = P_{ss,m-1} - P_{ttv,m-1}
\label{sq4}
\eeq
for $m \ge 1$, where the graph $G_{sq,e=ttv,m=1}$ is shown in
Fig.\ \ref{figsqexample} (e), and 
the $m=0$ graph of this type is taken to be the 5-vertex one at the right, with
\beq
P_{ttv,0}=q(q-1)(q^3-5q^2+10q-7)
\label{pttv0}
\eeq
Deleting one of the oblique bonds at the right end in Fig.\ 
\ref{figsqexample}(e) gives the relation 
\beq
P_{ttv,m} = D_4(q)P_{ss,m-1} - (q-2)P_{t,m}
\label{sq5}
\eeq
for $m \ge 1$.  Evidently, these linear relations have now closed, {\it i.e.},
no new
types of strip graphs are introduced beyond the $ss, st, tt, t, ttv$ that we
have.  Next, we substitute the assumed form (\ref{gamma}) into 
these equations.  Note that we do not have to specify $j_{max}$ or $k_{max}$; 
these are determined automatically by our solution of the equations.  This
substitution yields five linear equations for the five generating
functions for the different types of strips: the original one, $\Gamma_{ss}$,
together with the other four that have been generated by the operations of the
deletion-contraction theorem.  This set of equations can be written as the
linear transformation 
\beq
T v_{_{\Gamma}} = v_{_{0}}
\label{tv}
\eeq
where the vectors $v_{_{\Gamma}}$ and $v_{_{0}}$ are defined as 
\beq
v_{_{\Gamma}} = \left (\begin{array}{c}
                   \Gamma_{ss}  \\
                   \Gamma_{st}  \\
                   \Gamma_{tt}  \\
                   \Gamma_{t}   \\
                   \Gamma_{ttv} \end{array}\right )
\label{v}
\eeq
\beq
v_{_{0}} = \left (\begin{array}{c}
              P_{ss,0}+P_{st,0}   \\
              P_{st,0}+P_{tt,0}   \\
              P_{tt,0}+P_{t,0}    \\
              P_{t,0}                  \\
              P_{ttv,0}+(q-2)P_{t,0} \end{array}\right )
\label{v0}
\eeq
and, with this ordering of the basis, the matrix $T$ is given by 
\beq
T =  \left (\begin{array}{ccccc}
 1-(q-1)D_4(q)x & 1    & 0     & 0     & 0     \\
 -(q-1)(q-2)x   & 1    & 1     & 0     & 0     \\
 -(q-2)x        & 0    & 1     & 1     & 0     \\
  -x            & 0    & 0     & 1     & x     \\
 -D_4(q)x       & 0    & 0     & (q-2) & 1 \end{array}\right )
\label{tmatrix}
\eeq
The solution to eq.\ (\ref{tv}), $v_{_{\Gamma}} = T^{-1}v_{_{0}}$, 
yields the
desired generating function $\Gamma_{ss}$, together with generating functions
for the four other square strips of width 2 and different subgraphs on the 
right-hand ends.  In particular, all of the five generating functions in 
$v_{_{\Gamma}}$ have the same denominator ${\cal D}$ given as the determinant 
\beq
{\cal D} = \det(T)
\label{ddet}
\eeq
which is a general relation; in the present case, 
\beq
{\cal D} = 1 + \sum_{k=1}^2b_{sq(3),k} x^k
\label{ddetsum}
\eeq
where
\beq
b_{sq(3),1} = -(q-2)(q^2-3q+5)
\label{b1sqw2}
\eeq
and
\beq
b_{sq(3),2} = (q-1)(q^3-6q^2+13q-11)
\label{b2sqw2}
\eeq
Note the general property that
\beq
\det(T(x=0)) = 1
\label{tx0}
\eeq
whence 
\beq
{\cal D}(x=0) = 1
\label{dx0}
\eeq
which was already incorporated into our form (\ref{gammagen})-(\ref{d}) 
with $b_0=1$. 
For all of the five strips, we get $j_{max}=2$, {\it i.e.}, the respective 
generating functions have the form
\beq
\Gamma(G_{s(L_y);(e)},q,x) = \frac{a_{s(L_y);(e),0} + 
a_{s(L_y);(e),1}x}{1+b_{s(L_y),1}x+b_{s(L_y),2}x^2}
\label{gammamin}
\eeq
where, as above, the subscript $(e)$ denotes the initial (end) subgraph and is
implicit for the original strip, since its definition specifies the initial
subgraph $I$.  For brevity of notation, we shall sometimes also suppress the 
$(L_y)$ subscript, understanding it to be part of the type $s$. 
For the original square strip itself, the terms in the numerator polynomial 
${\cal N}$ are 
\beq
a_{sq(3),0} = q(q-1)D_4(q)^2
\label{a0sqw2}
\eeq
\beq
a_{sq(3),1} = -q(q-1)^3(q^3-6q^2+13q-11)
\label{a1sqw2}
\eeq
 From the results above, it is straightforward to obtain the numerators for the
other four generating functions of the strips with different subgraphs at the 
right.  For example, the numerator for $\Gamma_{tt} \equiv 
\Gamma(G_{sq(L_y=3,e=tt)},q,x)$ has 
\beq
a_{sq(L_y=3,e=tt),0} = P_{tt,0} = q(q-1)(q-2)^2
\label{a0tt}
\eeq
\beq
a_{sq(L_y=3,e=tt),1} = -q(q-1)^2(q-2)(q-3)
\label{a1tt}
\eeq
In Appendix 1 we present two alternate ways of calculating the generating
function $\Gamma(G_s,q,x)$ that are complementary to the method given above. 

\subsection{Some Basic Properties of Generating Functions}

   In this subsection we prove some basic properties of generating
functions using elementary properties of chromatic polynomials of arbitrary
graphs.  First, we observe that 
the zeroth order term in eq.\ (\ref{gamma}) is just the chromatic 
polynomial of the initial subgraph $I$: 
\beq
\Gamma(G_s,q,0) = {\cal N}(G_s,q,0) = a_{s,0} = P(I,q) 
\label{piq}
\eeq
Next, we have 

\vspace{3mm}

\begin{flushleft}
Theorem 2
\end{flushleft}

\beq
\Gamma(G_s,q,x) = 0 \  \quad {\rm for} \quad q=0,1,...\ell-1 \quad 
{\rm if} \quad G_s \supseteq K_\ell
\label{gammaqkell}
\eeq
whence
\beq
{\cal N}(G_s,q,x)=0  \ \quad {\rm for} \quad q=0,1,...\ell-1 \quad 
{\rm if} \quad G_s \supseteq K_\ell 
\label{nqkell}
\eeq
\beq
a_{s,j}(q) = 0  \ \quad {\rm for} \quad q=0,1,...\ell-1 \quad 
\forall \ j \quad {\rm if} \quad  G_s \supseteq K_\ell 
\label{aqkell}
\eeq
where $K_\ell$ denotes the complete graph with $\ell$ vertices. 

\vspace{3mm}

\begin{flushleft}
Proof
\end{flushleft}

    One requires at least $\ell$ colors to color a complete graph $K_\ell$
subject to the condition that no adjacent vertices have the same color (all
vertices are adjacent on a complete graph), {\it i.e.}, $P(K_\ell,q)=0$ for 
$q=0,1,...\ell-1$.  It follows that if $G$ is an arbitrary graph that 
contains at least one complete graph $K_\ell$, then $P(G,q)=0$ for
$q=0,1,...,\ell-1$.  In particular, this applies for the special case $G=G_s$. 
\ $\Box$. 

\vspace{2mm}

Since $K_1$ is just a vertex, and $K_2$ is just a bond, and since all $G_s$
graphs have bonds (indeed have no disjoint vertices), a general consequence of
Theorem 2 is that 
\beq
\Gamma(G_s,q,x) = 0 \ \quad {\rm for} \quad q=0, 1
\label{gammaq01}
\eeq
whence 
\beq
{\cal N}(G_s,q,x)=0 \ \quad {\rm for} \quad q=0, 1
\label{nq01}
\eeq
\beq
a_{s,j}(q) = 0  \ , \quad {\rm for} \quad q=0, 1 \quad \forall \ j
\label{aq01}
\eeq
The case $\ell=3$ in eqs. (\ref{gammaqkell})-(\ref{aqkell}), where the 
subgraph is a triangle, $K_3$, applies to our strip graphs of the triangular, 
kagom\'e, and $(3 \cdot 12^2)$ lattices.

   Another general result follows from the fact that an elementary upper 
bound on a chromatic polynomial is $P(G,q) \le q^n$ (for integer $q$)
since the right-hand side is the number of ways that one can color the
$n$-vertex graph $G$ without any constraint.  Indeed, this is what motivates
the use of the $1/n$'th power in the definition of $W(\{G\},q)$ in eq.\ 
(\ref{gs}).  Since $P(G,q)$ is a polynomial, this bound also applies for 
$q$ generalized from integer to real and complex values.  It follows that 
$W(\{G\},q)$ is bounded above by $W(\{G\},q) \le q$ for real $q$, and, more
generally, $|W(\{G\},q)| \le |q|$ for complex $q$.  It is thus convenient to
define, as we have in our earlier work \cite{w,w3,wa}, the reduced function 
\beq
W_r(\{G\},q) = q^{-1}W(\{G\},q)
\label{wr}
\eeq
which has a finite limit as $|q| \to \infty$.  In the case of regular lattices
$\{G\}=\Lambda$, in the context of large-$q$ Taylor expansions of reduced $W$
functions \cite{series,w,w3} it is also useful to define a related 
reduced function 
\beq
\overline W(\Lambda,q) = \frac{W(\Lambda,q)}{q(1-q^{-1})^{\zeta/2}} 
\label{wbar}
\eeq
where $\zeta$ is the coordination number of the lattice (in graph theory
terminology, the degree $\Delta$ of a vertex on the lattice).  The large-$q$
series expansion in terms of the variable 
\beq
y = \frac{1}{q-1}
\label{y}
\eeq
is then 
\beq
\overline W(\Lambda,y)=1+\sum_{n=1}^\infty w_n y^n
\label{wseries}
\eeq
(This assumes that $\overline W(\Lambda,y)$ is analytic at $y=1/q=0$; for a
recent discussion of this, see Refs. \cite{w,wa}.) 
We shall discuss these large-$q$ series below in connection with our exact 
solutions for $W(\{G_s\},q)$ for various strip graphs.

For strip graphs $(G_{s,kp})_m$ where $G_s$ is $k$-partite ($kp$), 
{\it i.e.}, can 
be uniquely colored with $k$ colors such that no two adjacent vertices have
the same color, the generating function satisfies 
\beq
\Gamma(G_{s,kp},q=k,x) = \frac{k!}{1-x}
\label{gammakpartite}
\eeq
reflecting the fact that
\beq
P((G_{s,kp})_m,k)=k!
\label{pgkpartite}
\eeq

\section{Results for Explicit Generating Functions}

   In this section we give results of our calculation of explicit generating 
functions for various types of strips.  Since some of these, in particular,
those associated with heteropolygonal Archimedean lattices, involve repeating
subgraph units $H$ more complicated than just single polygons, it is useful to
give formulas for the number of vertices as a function of the length of the 
strip, as measured by the number, $m$, of repeating subunits added to the 
initial subgraph $I$ forming the end of the strip.  We do this in Tables I and
II.  These tables also contain information on the degrees of the polynomials
$a_{s,j}$ and $b_{s,k}$ in the numerator and denominator of the generating
functions for various types of strip graphs.  As was evident in our proof
of Theorem 1 and will be discussed further in section IV (see Corollaries 
3.1 and 3.2), the choice of the initial subgraph $I$ affects only ${\cal N}$
and not ${\cal D}$ in the generating function and consequently does not affect
the asymptotic limiting function $W(\{G_s\},q)$.

\begin{table}
\caption{\footnotesize{
Properties of generating functions for strip graphs $(G_s)_m$ as 
defined in eq.\ (\ref{gs}) of width $L_y$, comprised of $m$ repetitions of a 
subgraph $H$, with an initial subgraph $I$, as discussed in the text, and 
having $j_{max}=1$ and $k_{max}=2$.  See the text for the definition of the 
diagonal width $d$.   
The degrees of $a_{s(L_y),j}$ and $b_{s(L_y),k}$, as polynomials in $q$, 
are denoted $deg(a_{s(L_y),j})$ and $deg(b_{s(L_y),k})$, and the 
subscript $s(L_y)$ is omitted to save space. $R_s$ is the discriminant defined
in eq.\ \ref{rs}.  $N_{b.p.}$ is the number of 
branch point singularities in the degeneracy equation (\ref{degeneracy}).
$N_{arcs}$ and $N_{s.c. arcs}$ denote the number of arcs and self-conjugate 
arcs, respectively, in the continuous locus of points ${\cal B}$ 
where $W(\{G_s\},q)$ is nonanalytic.}}
\begin{center}
\begin{tabular}{cccccccccc}
$G_s$ & $c_{s,1}$ & $c_{s,0}=deg(a_0)$ & $deg(a_1)$ & $deg(b_1)$ & $deg(b_2)$ 
& $deg(R_s)$ & $N_{b.p.}$ & $N_{arcs}$ & $N_{s.c. arcs}$ \\
sq, $L_y=3$        & 3 & 6 & 7 & 3 & 4 & 6  & 6  & 3 & 1 \\
sq$_d$, $d=3$      & 5 & 4 & 5 & 5 & 6 & 10 & 10 & 5 & 1 \\ 
tri, $L_y=3$       & 3 & 6 & 7 & 3 & 4 & 6  & 6  & 3 & 1 \\
$(3 \cdot 6 \cdot 3 \cdot 6)$,$L_y=2$ 
                   & 5 & 6 & 7 & 5 & 6 & 10 & 8  & 4 & 0 \\
$(3 \cdot 12^2)$, $L_y=2$ 
                   & 10 & 12 & 14 & 10 & 12 & 20 & 18 & 9 & 1 \\
$(4 \cdot 8^2)$, $L_y=2$ 
                   & 8  & 8  & 10 & 8  & 10 & 16 & 16 & 8 & 0 \\
\hline
\end{tabular}
\end{center}
\label{archtable}
\end{table}

\begin{table}
\caption{\footnotesize{
Properties of generating functions for strip graphs $(G_s)_m$ with 
$j_{max} > 1$, $k_{max} > 2$. Notation is the same as in Table 1.}}
\begin{center}
\begin{tabular}{cccccccccccc}
$s$ & $j_{max}$ & $k_{max}$ & $c_{s,1}$ & $c_{s,0}=deg(a_0)$ & $deg(a_1)$ & 
$deg(a_2)$ & $deg(a_3)$ & $deg(b_1)$ & $deg(b_2)$ & $deg(b_3)$ & $deg(b_4)$ \\
sq, $L_y=4$     & 2 & 3 & 4 & 8  & 10 & 12 & -  & 4 & 6 & 8  & - \\
hc, $L_y=3$     & 2 & 3 & 6 & 10 & 12 & 10 & -  & 6 & 8 & 6  & - \\
$sq_d$, $d=4$   & 2 & 3 & 6 & 8  & 10 & 12 & -  & 6 & 8 & 10 & - \\
tri, $L_y=4$    & 3 & 4 & 4 & 8  & 10 & 12 & 14 & 4 & 6 & 8  & 10 \\
$(4 \cdot 8^2)$ $L_y=3$ 
                & 2 & 3 & 12 & 16 & 20 & 18 & -  & 12 & 16 & 18  & - \\
\hline
\end{tabular}
\end{center}
\label{widertable}
\end{table}

  We next list the polynomials $a_{s,j}$ and $b_{s,k}$ that we have calculated
for several strip graphs $G_s$.  Others are listed in Appendix 2. 

\subsection{Strip of the Square Lattice}

   For strips of the square lattice of width $L_y$, we take the repeating
subgraph unit to be a vertical stack of $w=L_y-1$ squares, and the initial 
subgraph $I$ on the right to be identical to this repeating unit, $I=H$.  In
illustrating the proof of Theorem 1, we have already discussed in detail our 
calculation of the generating function for the square strip of width
$L_y=3$ and have listed the resultant polynomials 
$a_{sq(3),0}$, $a_{sq(3),1}$, 
$b_{sq(3),1}$, and $b_{sq(3),2}$ as eqs. (\ref{a0sqw2}), (\ref{a1sqw2}), 
(\ref{b1sqw2}), and (\ref{b2sqw2}).  
For the square strip of width $L_y=4$, we find $j_{max}=2$, $k_{max}=3$, and,
(with the notation $G_{sq(L_y=4)} \equiv G_{sq(4)}$) 
\beq
a_{sq(4),0} = q(q-1)D_4(q)^3
\label{a0sqw3}
\eeq
\beq
a_{sq(4),1} = -q(q-1)^3(q^2-4q+5)(2q^4-15q^3+45q^2-65q+40)
\label{a1sqw3}
\eeq
\beqs
a_{sq(4),2} & = & q(q-1)^3(q^8-16q^7+112q^6-449q^5+ \nonumber \\
&  & 1130q^4-1829q^3+1858q^2-1084q+279)
\label{a2sqw3}
\eeqs
\beq
b_{sq(4),1} = -q^4+7q^3-23q^2+41q-33
\label{b1sqw3}
\eeq
\beq
b_{sq(4),2} = 2q^6-23q^5+116q^4-329q^3+553q^2-517q+207
\label{b2sqw3}
\eeq
\beq
b_{sq(4),3} = -q^8+16q^7-112q^6+449q^5-1130q^4+1829q^3-1858q^2+1084q-279
\label{b3sqw3}
\eeq

\subsection{Strip of the Triangular Lattice}

   We consider a strip of the triangular lattice of width $L_y$. The $H$ and
$I$ subgraphs were defined above in section II. 
The strip graph of this type with $m$ repeated $H$ subgraph 
units is denoted $(G_{t(L_y)})_m$.  
The case $L_y=2$ yields a chromatic polynomial
that factorizes, as in eq.\ (\ref{pghfactor}).  For $L_y=3$ we find 
$j_{max}=1$ and $k_{max}=2$ with 
\beq
a_{t(3),0} = q(q-1)(q-2)^4
\label{a0triw2}
\eeq
\beq
a_{t(3),1} = -q(q-1)^2(q-2)^3(q-3)
\label{a1triw2}
\eeq
\beq
b_{t(3),1} = -q^3+7q^2-18q+17
\label{b1triw2}
\eeq
\beq
b_{t(3),2} = (q-2)^3(q-3)
\label{b2triw2}
\eeq
The results for the next wider strip, with $L_y=4$, are somewhat lengthy and
are given in Appendix 2.  

\subsection{Strip of the Honeycomb Lattice}

    The minimal-width strip for the honeycomb (also called hexagonal) 
lattice with a nonfactorizing chromatic polynomial is shown in 
Fig.\ \ref{figstrip}(e) and has a width $L_y=3$ (represented as a brick
lattice). In contrast to the strips of the square and 
triangular lattice with the same $L_y=3$ width, 
the honeycomb strip yields a generating function with numerator and 
denominator functions that are of higher degrees in $x$, {\it viz.}, 
$j_{max}=2$ and $k_{max}=3$.  We find 
\beq
a_{hc(3),0} = q(q-1)D_6(q)^2
\label{a0hcw2}
\eeq
\beqs
a_{hc(3),1} & = & -q(q-1)^2(q^9-13q^8+76q^7-262q^6+589q^5 \nonumber \\
              &  & -901q^4+943q^3-651q^2+266q-47)
\label{a1hcw2}
\eeqs
\beq
a_{hc(3),2} = q(q-1)^7(q-2)^2
\label{a2hcw2}
\eeq
\beq
b_{hc(3),1} = -q^6+8q^5-28q^4+56q^3-71q^2+58q-26
\label{b1hcw2}
\eeq
\beq
b_{hc(3),2} = (q-1)^2(q^6-10q^5+43q^4-102q^3+144q^2-120q+49)
\label{b2hcw2}
\eeq
\beq
b_{hc(3),3} = -(q-1)^4(q-2)^2
\label{b3hcw3}
\eeq

\subsection{Strip of the Kagom\'e Lattice}

   It is also of interest to calculate generating functions for the chromatic
polynomials of strips of lattices which involve more than just one regular
polygons.  The uniform lattices that involve tilings of the plane by one or
more regular polygons are called Archimedean (for a review, see
Ref. \cite{gs}); for some related studies, see Refs. \cite{w3,wn,cmo}) 
and, since each vertex is equivalent to every other, these lattices may be 
defined by the sequence of polygons that one traverses when making a circuit 
(say, in a counterclockwise direction) around any vertex, 
\beq
\Lambda = \prod_i p_i^{a_i}
\label{lambda}
\eeq
where $p_i$ label the polygon and $a_i \ge 1$ is the number of times that a
given polygon occurs contiguously in the product.  The total number of times
that a polygon of type $p_i$ occurs is the sum of its contiguous and
noncontiguous appearances.   The Archimedean lattices include three
homopolygonal (also called monohedral) cases: square, triangular, and
honeycomb.  The remaining Archimedean lattices are heteropolygonal, 
{\it i.e.}, are composed of more than one type of regular polygon. 
For the strips of heteropolygonal lattices, it is
necessary to fix a convention to describe the width.  We shall choose the
convention of taking the width to be the number of layers of the polygon with 
the greatest number of sides, and, to retain a correspondence with the
homopolygonal lattices, we shall label the polynomials in $\Gamma(G_s,q,x)$ by
$w+1$ (which is $L_y$ for the homopolygonal strips). 
For our present study we consider first a 
strip of the kagom\'e lattice, denoted $(3 \cdot 6 \cdot 3 \cdot 6)$ in the
standard mathematical notation (\ref{lambda}).  We take the strip to be 
of width $w=1$ and oriented as shown in Fig.\ \ref{figstrip}(f).
The initial subgraph $I$ on the right is a
hexagon, and the repeating subgraph unit $H$ is the graph comprised of a
hexagon and the two triangles adjacent to it.  We find $j_{max}=1$,
$k_{max}=2$ and 
\beq
a_{kag(2),0} = q(q-1)D_6(q)
\label{a0kag}
\eeq
\beq
a_{kag(2),1} = -q(q-1)^5(q-2)
\label{a1kag}
\eeq
\beq
b_{kag(2),1} = -(q-2)(q^4 - 6q^3 + 14q^2 -16q + 10)
\label{b1kag}
\eeq
\beq
b_{kag(2),2} = (q-1)^3(q-2)^3
\label{b2kag}
\eeq

   We have also calculated generating functions for two other heteropolygonal
Archimedean lattices, {\it viz.}, the $(3 \cdot 12^2)$ and $(4 \cdot 8^2)$
lattices.  The results for these are given in Appendix 2.  We next proceed to
study the limit in which the length of a strip goes to infinity. 

\section{Analytic Structure of $W(\{ G_{\lowercase{s}} \},\lowercase{q})$}

   To calculate $W(\{G_s \},q)$ from the generating function $\Gamma(G_s,q,x)$,
we prove a basic theorem.  Let us write the denominator function ${\cal D}$
of the generating function $\Gamma(G_s,q,x)$ in factorized form, in terms of
its roots $r_{s,k}$, $k=1,...,k_{max}$, as 
\beq
{\cal D}(G_s,q,x) = 1 + \sum_{k=1}^{k_{max}} b_{s,k}(q) x^k = 
b_{s,k_{max}}\prod_{k=1}^{k_{max}}(x-r_{s,k}(q))
\label{dfactor}
\eeq
or equivalently 
\beq
{\cal D}(G_s,q,x) = \prod_{k=1}^{k_{max}}(1-\lambda_{s,k}(q)x)
\label{lambdaform}
\eeq
where 
\beq
\lambda_{s,k}(q)=\frac{1}{r_{s,k}(q)}
\label{lambdaroot}
\eeq
(and the roots satisfy the relation 
$b_{s,k_{max}}(-1)^{k_{max}}\prod_{k=1}^{k_{max}}r_{s,k} = 1$.) 

\vspace{4mm}

\begin{flushleft}
Theorem 3 
\end{flushleft}

   Consider a strip graph $G_s$ as defined in eq.\ (\ref{gs}), constructed by 
$m$ successive additions of a repeating graphical subunit $H$ to an initial 
subgraph $I$.  For a given value of $q$, denote the function $\lambda_{s,k}(q)$
of maximal magnitude in eq.\ (\ref{lambdaform}) as $\lambda_{s,max}(q)$. 
Then the $n \to \infty$ asymptotic limiting function $W(\{ G_s \},q)$ as 
defined in eq.\ (\ref{w}) with eq.\ (\ref{wdefqn}) is given by 
\beq
W(\{G_s \},q) = (\lambda_{s,max}(q))^{1/c_{s,1}}
\label{wlambda}
\eeq
where the definition of $c_{s,1}$ was given in eq.\ (\ref{nvertices}) and a
general formula for it was presented in eq.\ (\ref{cs1}).  The phase to pick in
evaluating eq.\ (\ref{pphase}) and eq.\ (\ref{wlambda}) is chosen so that for
sufficiently large real $q$, $W(\{G_s \},q)$ is real.

\vspace{3mm} 

\begin{flushleft}
Proof
\end{flushleft}

   First, for the case $k_{max}=1$, where the chromatic polynomial for the
strip factorizes according to eq.\ (\ref{pghfactor}), one recognizes 
\beq
\lambda_s(q) = -b_{s,1}(q) = P_r(H,q)
\label{lambdasone}
\eeq
where $P_r(H,q)$ was given in eq.\ (\ref{phkell}), and the result
(\ref{wlambda}) then follows immediately from the elementary expansion 
\beq
\Gamma(G_s,q,x) = 
\frac{{\cal N}(G_s,q,x)}{1-\lambda_s(q)x} = 
{\cal N}(G_s,q,x)\sum_{m=0}^\infty \lambda_s(q)^m x^m
\label{onelambda}
\eeq
together with eqs. (\ref{gamma}) and (\ref{w}), whence 
\beq
W(\{G_s\},q) = \lim_{n \to \infty} (\lambda_s(q))^{m/n} = 
(\lambda_s(q))^{1/c_{s,1}}
\label{wonelambda}
\eeq
where we have used eq.\ (\ref{nvertices}).  Note that, for the generic case
where $c_{s,1} \ge 2$, $W(\{G_s\},q)$ has branch point singularities at the
zeros (if any) of $\lambda_s(q)$.   We shall next 
discuss the case $k_{max}=2$ and comment on the straightforward generalization
to $k_{max} \ge 3$. For brevity of notation, we shall suppress
the $q$-dependence of the $\lambda_{s,k}(q)$.  For $k_{max}=2$, eq.\ 
(\ref{gammagen}) 
becomes
\beqs
\Gamma(G_s,q,x) & = & \frac{{\cal N}(G_s,q,x)}
{(1-\lambda_{s,1}x)(1-\lambda_{s,2}x)}
= \frac{{\cal N}(G_s,q,x)}{(\lambda_{s,1} - \lambda_{s,2})}
\biggl [ \frac{\lambda_{s,1}}{1-\lambda_{s,1}x} -
 \frac{\lambda_{s,2}}{1-\lambda_{s,2}x} \biggr ] \nonumber \\
& & \nonumber \\
& = & \frac{{\cal N}(G_s,q,x)}{(\lambda_{s,1}-\lambda_{s,2})}
\sum_{m=0}^\infty \Bigl ( \lambda_{s,1}^{m+1}-\lambda_{s,2}^{m+1}\Bigr )x^m
\label{keq2}
\eeqs
(which is, of course, symmetric under $\lambda_{s,1} \leftrightarrow 
\lambda_{s,2}$.)
Then, from the definition (\ref{w}) with (\ref{wdefqn}), taking the $1/n$'th
power as $n \to \infty$, and again using eq.\ (\ref{nvertices}), we obtain 
the result (\ref{wlambda}).  It is straightforward to generalize this to the 
case $k_{max} \ge 3$.  In passing, we
note that if some fixed subset $N_{eq} \ge 2$ of the $\lambda_{s,k}$ were 
identically equal, this would not affect the result (\ref{wlambda}).  (This is
similar to the situation where some number of the eigenvalues of a transfer 
matrix for a spin model are identically equal \cite{z6}.)  This is trivial if
the identical $\lambda_{s,k}$'s are not maximal; if they are, it is also clear
from the fact that
\beq
\lim_{n \to \infty} \Bigl ( N_{eq} (\lambda_{s,max}(q))^{m+1} \Bigr )^{1/n} 
= (\lambda_{s,max}(q))^{1/c_{s,1}} \ , 
\label{neq}
\eeq
independent of $N_{eq}$.  (This possibility is mentioned only for 
completeness; we have not encountered any strip graph $G_s$ where any 
$\lambda_{s,k}(q)$ are identically equal.)   The choice of phase, $r=0$ in
eq.\ (\ref{pphase}) is determined by the fact that for sufficiently large real 
$q$, $\lambda_{s,max}(q) > 0$. 

This completes the proof. \ $\Box$

\vspace{3mm}

   We next state some corollaries to this theorem. 

\vspace{2mm}

\begin{flushleft}
Corollary 3.1
\end{flushleft}

  Given the strip graph $G_s$ as defined in eq.\ (\ref{gs}), the asymptotic
function $W(\{G_s\},q)$ as defined in (\ref{w}) with (\ref{wdefqn}) is 
independent of the numerator function ${\cal N}(G_s,q,x)$.  It follows also, 
{\it a fortiori}, that the continuous locus of points ${\cal B}$ where 
$W(\{ G_s \},q)$ is nonanalytic, is independent of ${\cal N}(G_s,q,x)$. 

\vspace{3mm}

\begin{flushleft}
Corollary 3.2
\end{flushleft}

  Given the strip graph $G_s$ as defined in eq.\ (\ref{gs}), the asymptotic
function $W(\{G_s\},q)$ as defined in (\ref{w}) with (\ref{wdefqn}) is
independent of the initial subgraph $I$ and depends only on the infinitely
repeated subgraph unit $H$.  Furthermore, {\it a fortiori}, the continuous 
locus of points 
${\cal B}$ where $W(\{G_s\},q)$ is nonanalytic is independent of $I$ and
depends only on $H$. 

\vspace{3mm}

In particular, for this second corollary, recall that our proof of Theorem 1 
showed that ${\cal D}$ was independent of the initial subgraphs of the strip 
graphs generated by the operation of the contraction-deletion theorem, as was
noted before eq.\ (\ref{ddet}).  
It is interesting to see Corollaries 3.1 and 3.2 in operation by
studying the chromatic zeros of strip graphs of a given type $G_s$, varying the
type of initial subgraph $I$ and the number of repeated subgraph units $m$.
For small $m$, we find, as expected, that the chromatic zeros clearly depend 
on what $I$ one uses, but as $m$ gets larger and larger, this dependence is 
progressively reduced, until it disappears altogether as $m \to \infty$. 
Moreover, as we shall discuss in later work
\cite{strip2}, for strip graphs of the form $J(\prod_{\ell=1}^m H)I$, the 
asymptotic $m \to \infty$ limiting functions $W(\{ G_s \},q)$ do, in general, 
depend on both of the end graphs $I$ and $J$. 

\vspace{2mm}

   For a strip graph $(G_s)_m$ composed of $m$ repeated subgraph units $H$
adjoined to an initial subgraph $I=H$ and having the property that the 
chromatic polynomial factorizes as in (\ref{pghfactor}), the asymptotic 
limiting function is given by 
\beq
W(\{G_{s;K_\ell}\},q) = (P_r(H,q))^{1/(n(H)-\ell)}
\label{wkmax1}
\eeq
where we have used eq.\ (\ref{cs1}) for $c_{s,1}$.  For the proof of this
result, we observe that from the formula (\ref{gammakell}), one can 
immediately identify $\lambda_s(q) = P_r(H,q)$, and eq.\ (\ref{wkmax1}) then
follows from Theorem 3 and its special case eq.\ (\ref{wonelambda}) for 
$k_{max}=1$.  For example, for strips of the square and triangular lattices of 
width $L_y=2$, using the $\ell=2$ special case of eq. (\ref{wkmax1}),
we obtain 
\beq
W(\{G_{sq(2)}\},q)=(q^2-3q+3)^{1/2}
\label{wsqw1}
\eeq
and
\beq
W(\{G_{t(2)}\},q)=q-2
\label{wtriw1}
\eeq

  For the case $k_{max}=2$, the roots of ${\cal D}(G_s,q,x)=0$ as an equation
in $x$ are $r_{s,k}$, $k=1,2$, 
given by $r_{s,\pm} = (-b_{s,1} \pm \sqrt{R_s})/(2b_{s,2})$, so that 
the $\lambda_{s,k}$, $k=1,2$, are
\beq
\lambda_{s,\pm} = r_{s,\pm}^{-1} = -\frac{1}{2}\Bigl [ b_{s,1} \pm \sqrt{R_s}
\ \Bigr ]
\label{lambdak2}
\eeq
where the discriminant $R_s$ is 
\beq
R_s = b_{s,1}^2 - 4b_{s,2}
\label{rs}
\eeq

\vspace{3mm}

\begin{flushleft}
Corollary 3.3
\end{flushleft}

  Given the strip graph $G_s$ as defined in eq.\ (\ref{gs}), the maximal
$\lambda_{s,max}(q)$ in eqs. (\ref{lambdaform}) and (\ref{wlambda}) has the
behavior, for large real $q$, 
\beq
\lambda_{s,max}(q) \sim q^{c_{s,1}} \ \quad {\rm as} \quad q \to \infty
\label{lambdalargeq}
\eeq

\vspace{3mm}

\begin{flushleft}
Proof
\end{flushleft}

   A chromatic polynomial for an arbitrary $n$-vertex graph $G$ is of degree 
$n$ as a polynomial in $q$ (with highest-degree term is $q^n$).  Hence, from 
the 
definition (\ref{w}), it follows that $W(\{G\},q) \sim q$ as $q \to \infty$ 
for real $q$ (more generally, $|W(\{G\},q)| \sim |q|$ for complex $q$).
Combining this with Theorem 3, eq. (\ref{wlambda}) yields the result
(\ref{lambdalargeq}).  

\vspace{4mm}

   We next prove a theorem which explicitly determines the continuous locus of 
points ${\cal B}$ where $W(\{G_s\},q)$ is nonanalytic: 

\begin{flushleft}
Theorem 4
\end{flushleft}

  Let a strip graph $G_s$ be defined as in eq.\ (\ref{gs}).  Then the
continuous locus of points ${\cal B}$ where the asymptotic function 
$W(\{G_s\},q)$ is nonanalytic is comprised of the points where two (nonzero) 
$\lambda_{s,k}$ terms in eq.\ (\ref{lambdaform}) both become maximal and
degenerate in magnitude, {\it i.e.}, the solutions to the equation 
\beq
  \Big | \lambda_{s,max}(q) \Big | =  \Big | \lambda_{s,max'}(q) \Big |
\label{degeneracy}
\eeq

\begin{flushleft}
Proof
\end{flushleft}

   This follows from the general analysis of $W(\{G\},q)$, specifically part
(d) of Theorem 1, given in Ref. \cite{w}.  Note that, from eq.\ 
(\ref{wlambda}),
if $c_{s,1} \ge 2$, then $W(\{G_s\},q)$ has isolated branch point 
singularities where $\lambda_{s,max}(q)=0$, but, in general, these do not lie
on the continuous locus ${\cal B}$. \ $\Box$. 

   A simple corollary to this theorem is that if the strip graph $G_s$ 
defined by eq.\ (\ref{gs}) has the property that its
chromatic polynomial factorizes in the form of eq.\ (\ref{pghfactor}), then 
$W(\{G_s\},q)$, which we have calculated in eq.\ (\ref{wkmax1}), is analytic in
the complex $q$ plane, except for isolated branch points, so that 
${\cal B}$ is the null set, $\emptyset$.  This follows 
because in this case there is only one $\lambda_s$, (recall eq.\ 
(\ref{onelambda})) and hence no possibility for a degeneracy of maximal 
$\lambda_{s,k}$'s or resultant locus ${\cal B}$.
For example, in the case where $H$
is a $p$-sided polygon and each successive $p$-gon intersects the next on a
mutual edge so that $L_H=K_2$, we have 
\beq
W(\{G_{p-gons;K_2}\},q) = (D_p(q))^{1/(p-2)}
\label{wgpgons}
\eeq
where $D_p(q)$ was defined in eq.\ (\ref{dk}). The continuous locus 
${\cal B}=\emptyset$ for these graphs.  For $p \ge 4$, the 
function $W(\{G_{p-gons;K_2}\},q)$ has isolated branch point singularities 
at the $p-2$ zeros of the polynomial $D_p(q)$, but these are not connected 
with any continuous locus of singularities ${\cal B}$.  Note that $D_p(q)$ is a
polynomial of degree $p-2$; for odd $p \ge 3$, $D_p(q)$ has a single real 
zero at $q=2$ and $(p-1)/2$ conjugate pairs of complex zeros, while for 
even $q$, $D_p(q)$ has only complex zeros. 

   Before proceeding to discuss $W(\{G_s\},q)$ for specific strip graphs $G_s$,
we recall that the elementary property that for any family of graphs 
$\{G\}$, the associated locus of points ${\cal B}$ is invariant under complex 
conjugation:
\beq
{\cal B}(q) = {\cal B}(q^*)
\label{bcc}
\eeq
This follows from the fact that the coefficients of each power of $q$ in a 
chromatic polynomial $P(G,q)$ are real (indeed, integers) and the fact that 
${\cal B}$ originates as a merging of zeros of $P(G,q)$ as the number of
vertices $n \to \infty$ \cite{w}.  

\vspace{2mm}

\section{$W(\{G_{\lowercase{s}} \},\lowercase{q})$ for Specific Strip Graphs}

\begin{figure}
\begin{center}
\leavevmode
\epsfxsize=3.5in
\epsffile{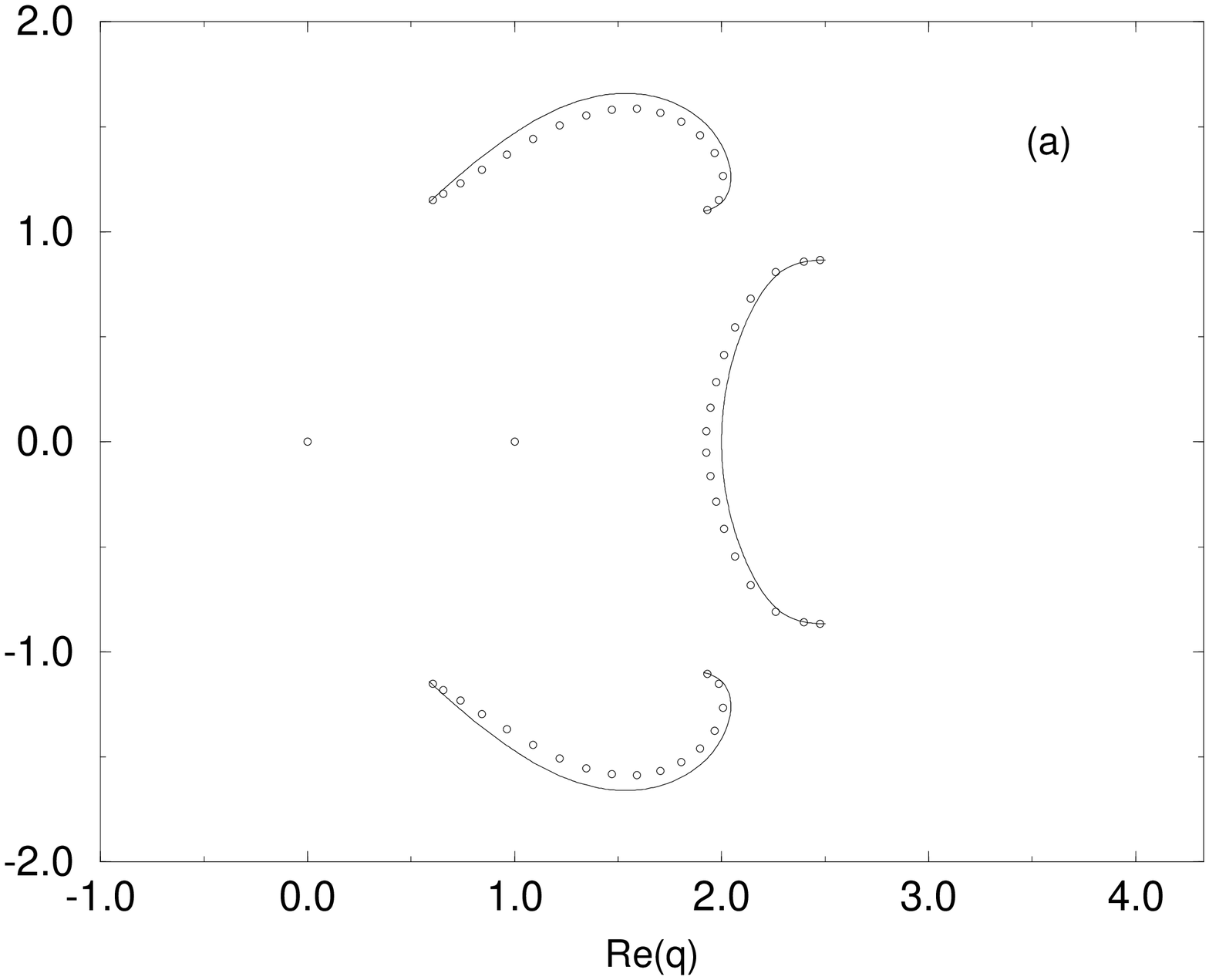}
\end{center}
\vspace{-4cm}
\begin{center}
\leavevmode
\epsfxsize=3.5in
\epsffile{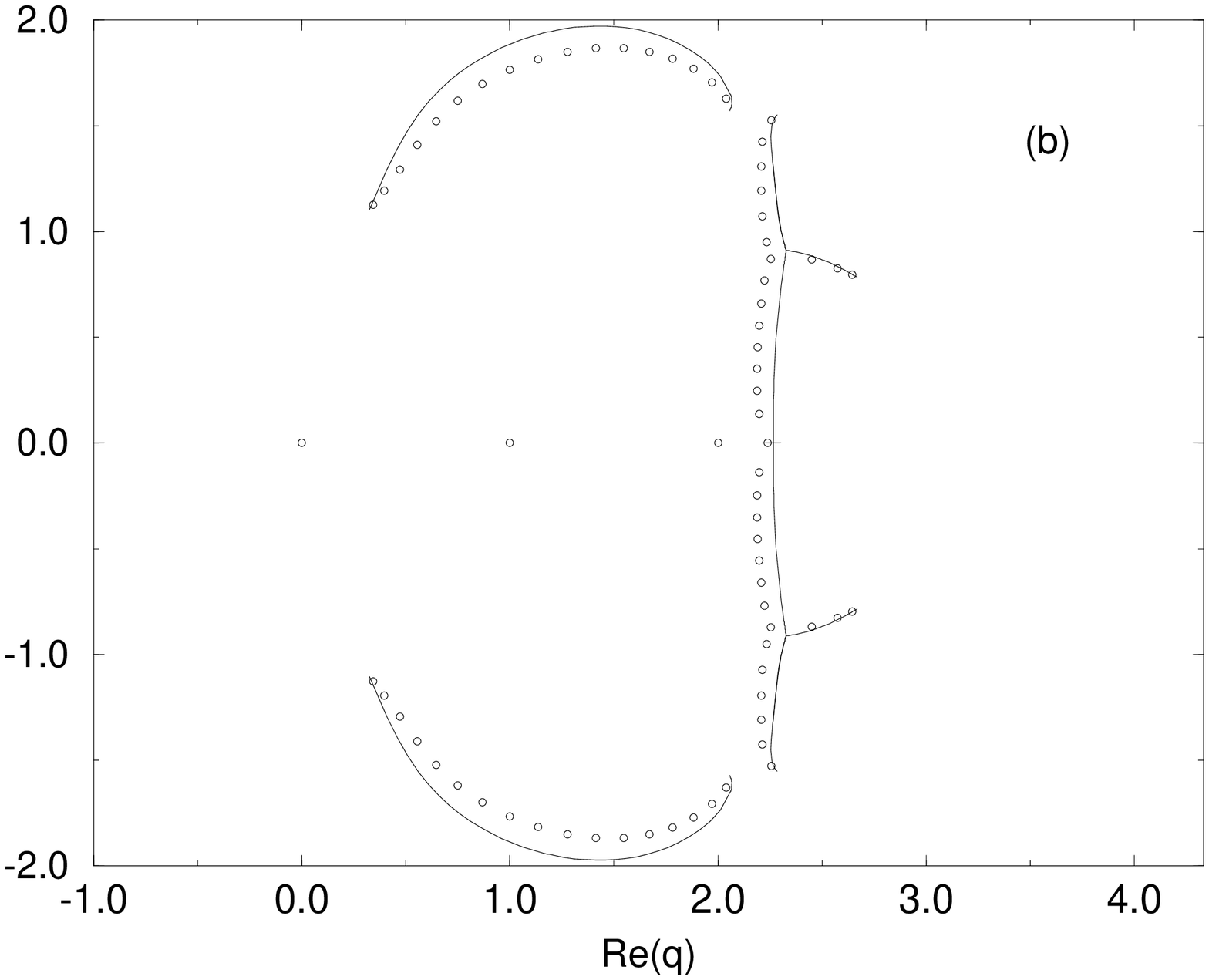}
\end{center}
\caption{\footnotesize{
Analytic structure of the function $W(\{G_{sq(L_y)}\},q)$ for 
$L_y =$ (a) 3, (b) 4, where $\{G_{sq(L_y)}\}$ denotes the 
$(L_x=\infty) \times L_y$ strip of the square lattice. 
The arcs are the continuous locus of points ${\cal B}$ where this function is 
nonanalytic. For comparison, the zeros of the chromatic polynomial
$P((G_{sq(L_y)})_m,q)$ for (a) $L_y=3$, $m=16$ (hence $n=54$ vertices) and 
(b) $L_y=4$, $m=16$ (hence $n=72$) are shown. In this and subsequent figures, 
the initial subgraph $I$ is specified in the text.}}
\label{figsq}
\end{figure}

   We next apply our general theorems to calculate the exact $W(\{G_s\},q)$ 
functions for specific infinitely long strip graphs.  The continuous locus of 
points ${\cal B}$ where $W(\{G_s\},q)$ is nonanalytic is shown in each of the 
accompanying figures.  As was discussed above, as the length of the strip graph
goes to infinity, the locus ${\cal B}$ forms by the merging together of the
zeros of the chromatic polynomial ({\it i.e.}, the chromatic zeros) of this 
strip graph.  One gains insight into how this limit is approached by 
calculating chromatic zeros for reasonably long finite strips of each type.  
Accordingly, we show these in the figures.  
There are also isolated chromatic zeros that
occur at $q=q_s=0,1$, and, for graphs that contain one or more triangles, at
$q_s=2$.  As discussed in Ref. \cite{w} and the Introduction, 
in cases (i) where these occur via factors that scale like the lattice size, 
{\it i.e.}, $(q-q_s)^{tn}$, $t \ge 1$, they remain as zeros of the limiting 
function $W(\{G\},q)$; (ii) if, on the other hand, they  
occur as factors like $(q-q_s)^u$ where $u$ does not scale like $n$, then they
do not appear in $W(\{G\},q)$.  (If one were to use the alternate definition 
based on the opposite order of limits in (\ref{wnoncomm}), then these zeros at 
$q=q_s$ would appear as discontinuities in $W(\{G_s\},q)$.) 
In case (ii) the different orders of limits in eq.\ (\ref{wnoncomm}) do not 
commute, whereas in case (i) they
do. Depending on the type of strip graph, isolated real chromatic zeros
may occur at certain other points also.  For example, in Fig.\ 
\ref{figsq}(b) 
there is a chromatic zero near to, but not precisely at, $q=2$ and another at
$q \simeq 2.239$, while in Fig.\ \ref{figtri}(b) there is a chromatic zero at 
$q=(3+\sqrt{5})/2 = 2.6180...$, 
etc.  The simplest types of strip graphs to consider are those with chromatic
polynomials that factorize in the manner of eq.\ (\ref{pghfactor}),
corresponding to $k_{max}=1$; we have
already discussed these above. 

\eject

    We proceed to consider strips graphs with denominator functions with
$k_{max}=2$, {\it i.e.}, ${\cal D}(x)$ that are quadratic in $x$ (listed in 
Table 1).  For these, we have the explicit results (\ref{lambdak2}), 
(\ref{rs}).  We first consider the strip graph of the square lattice of size 
$L_x \times L_y = \infty \times 3$.  (As was noted at the beginning, we always 
use free boundary conditions in this paper.)  From eqs. (\ref{b1sqw2}) and 
(\ref{b2sqw2}), we calculate the $W$ function by starting on the real axis 
with $q > 2$ and analytically continuing from there.  In this range, 
$b_{sq(3),1} < 0$ and so $\lambda_{sq(3),max}=\lambda_{sq(3),-}$, where 
$\lambda_{s,\pm}$ was given in eq. (\ref{lambdak2}).  Hence, in this range, 
\beqs
W(\{G_{sq(3)}\},q) & = & 2^{-1/3}\biggl [ (q-2)(q^2-3q+5) + 
\nonumber \\
& & \Bigl [(q^2-5q+7)(q^4-5q^3+11q^2-12q+8) \Bigr ]^{1/2} \biggr ]^{1/3}
\label{wsqw2}
\eeqs
In Fig.\ \ref{figsq}(a) we show our calculation of the exact continuous 
locus 
of points ${\cal B}$ where $W(\{G_{sq(3)}\},q)$ is nonanalytic. These consist 
of two complex-conjugate (c.c.) arcs together with a self-conjugate arc that
crosses the real axis at $q=2$, where $b_{sq(3),1}$ vanishes. Since ${\cal B}$ 
does not enclose any regions, we can analytically continue the 
expression (\ref{wsqw2}) throughout the $q$ plane, exclusive of the points on 
the locus ${\cal B}$ itself.  The two
complex zeros of $b_{sq(3),1}$ at $q=3/2 \pm (i/2)\sqrt{11}) = 
1.5 \pm 1.66i$ lie on the upper and lower arcs.  
The discriminant of the square root in eq.\ (\ref{wsqw2}) has zeros at 
$q=5/2 \pm (i/2)\sqrt{3}$ (for the quadratic factor)
and $q=0.587 \pm 1.14i$ and $1.91 \pm 1.10i$; (for the quartic factor).  These 
zeros are all simple and hence are branch points of the square root. 
As is evident in Fig.\ \ref{figsq}(a), the first pair form the endpoints of 
the self-conjugate arc crossing the real axis, while the second two pairs 
form the endpoints of the two complex-conjugate arcs.  In Fig.\ \ref{figsq}(a) 
we also show our calculations of the chromatic zeros for a long finite strip of
the square lattice\footnote{
We recall the well-known fact that calculations of the zeros of 
chromatic polynomials are difficult because for a graph $G$ with $n$ vertices,
the coefficients $p_\ell$ in the chromatic polynomial $P(G,q) = \sum_{\ell=1}^n
p_n q^n$ range from $p_n=1$ to much larger values for coefficients of terms 
$q^\ell$ with $\ell \simeq n/2$, as is clear from the theorem \cite{rtrev} that
$|p_r| \ge {{n-1}\choose {r-1}}$.  For this reason, it is necessary to use a
large number of digits of accuracy in routines to solve for the roots.}. 
As expected, aside from the generally occurring isolated
zeros at $q=0$ and 1, these zeros lie close to the arcs that comprise 
${\cal B}$ in the infinite-length limit.  

   Next, in Fig.\ \ref{figsq}(b) we present our exact calculation for 
${\cal B}$ for a square strip of the next greater width, {\it i.e.}, 
size $L_x \times L_y = \infty \times 4$, together with chromatic zeros 
computed for a long finite strip.  Our explicit analytic expression for 
$W(\{G_{sq(4)}\},q)$, which involves the root of a cubic equation, is rather 
complicated and hence is not given here; it can be obtained from eqs. 
(\ref{b1sqw3})-(\ref{b3sqw3}).   The comparison of the results presented in 
Fig.\ \ref{figsq}(b) with those for the $L_y=3$ strip in Fig.\ \ref{figsq}(a) 
shows how this locus ${\cal B}$ changes as the width of a strip of a
given type increases.  We see that all three of the arcs elongate, and the 
single self-conjugate arc of the $L_y=3$ strip develops a
more complicated structure for the $L_y=4$ strip.  In particular, multiple 
points appear on this arc for $L_y=4$.  Here, we use the term ``multiple 
point'' in the technical sense of algebraic geometry, meaning a point on an 
algebraic curve where two or more branches of the curve meet \cite{alg,cmo}.  
The index of the multiple point is the number of branches that meet; in Fig.\ 
\ref{figsq}(b) there is (i) a multiple point at $q \simeq 2.265$ where a 
vertical component of ${\cal B}$ intersects a small line segment in 
${\cal B}$ on the real axis, and (ii) two multiple points forming a 
complex-conjugate pair, at which three components of ${\cal B}$ meet.  
It is also interesting to compare the exact arcs that we have
calculated for infinite strips, and the associated chromatic zeros for finite 
strips, with the chromatic zeros calculated in Ref. \cite{baxter} for an 
$L_x \times L_y = 8 \times 8$ square lattice with the same (free) boundary 
conditions.  Already for a width $L_y=4$, the chromatic zeros for the strip 
resemble those of the larger patch of the 2D square lattice.  There are 
branchings in the chromatic zeros for the $8 \times 8$ lattice at 
complex-conjugate points close to the locations where we find multiple points 
on ${\cal B}$ for the $\infty \times 4$ strip.  Our previous results on 
$W(\{G\},q)$ in Ref. \cite{w} have shown that boundary conditions can 
strongly affect ${\cal B}$ and the associated chromatic zeros that merge to
form this locus in the $n \to \infty$ limit of an $n$-vertex graph.  For
example, ${\cal B} = \emptyset$ for the infinite limit of a 
linear graph with free boundary conditions, whereas for the same graph with
periodic boundary conditions, {\it i.e.}, the infinite-$n$ limit of the 
circuit graph, we found that ${\cal B}$ is the circle $|q-1|=1$, which 
separates the complex $q$ plane into two distinct regions\footnote{
In contrast, in statistical mechanics, the existence of the
thermodynamic limit entails independence of boundary conditions, except in the
well-understood sense that they can determine which of degenerate long range
orderings the system takes in a low-temperature, broken-symmetry phase. 
This is one of several differences between the behavior of $W(\Lambda,q)$ and 
the somewhat analogous function for a spin model in statistical mechanics, 
$\exp(f(\Lambda,K))$, where $f$ is the per site free energy on a lattice 
$\Lambda$, and $K=\beta J$ as in the text. Another is the noncommutativity of 
eq.\ (\ref{wnoncomm}), which has no analogue in statistical mechanics.  
See Ref. \cite{w} for further discussion.}. 
Nevertheless, it is intriguing to compare our results with the chromatic 
zeros computed for an $8 \times 8$ lattice with cylindrical boundary 
conditions, {\it i.e.}, periodic in one direction and free in the
other \cite{baxter}.  We observe that there are cusp-like structures in 
the chromatic zeros on the $8 \times 8$ lattice in the complex-conjugate 
regions where there are gaps between the arcs on the $\infty \times 4$ strip. 

   Using our results for $W(\{G_{sq(L_y)}\},q)$, 
we can calculate the large-$q$ series for the reduced functions 
$W_r(\{G_{sq(L_y)}\},q)$ defined for a general graph $G$ 
in eq.\ (\ref{wr}).   In order to compare with the large-$q$ series expansion
for the infinite square lattice, we actually consider the function 
$\overline W(\{G_{sq(L_y)}\},q)$ defined in eq. (\ref{wbar}).  (For this
purpose, we formally substitute $\zeta=4$, although not all of the vertices of
the strip graphs have the same degree.)  With $y=1/(q-1)$ as in eq. (\ref{y}), 
we find $\overline W(\{G_{sq(1)}\},q) = 1+y$, 
$\overline W(\{G_{sq(2)}\},q)=(1+y)(1-y+y^2)^{1/2}$, whence 
\beqs
\overline W(\{G_{sq(2)}\},q) & = & 1 + \frac{1}{2}y - \frac{1}{2^3}y^2 + 
\frac{9}{2^4}y^3 + \frac{27}{2^7}y^4 - \frac{9}{2^8}y^5 - 
\frac{117}{2^{10}}y^6 - \frac{135}{2^{11}}y^7 + O(y^8) \nonumber \\
& \simeq & 1 + 0.5y -0.125y^2 + 0.5625y^3 + 0.2109y^4 \nonumber \\ 
& & -0.0352y^5 - 0.1143y^6 - 0.06592y^7 + O(y^8) 
\label{wbarseriessqw1}
\eeqs
and 
\beqs
\overline W(\{G_{sq(3)}\},q) & = & 1 + \frac{1}{3}y - \frac{1}{3^2}y^2 
+ \frac{59}{3^4}y^3 + \frac{44}{3^5}y^4 - \frac{32}{3^6}y^5 
- \frac{613}{3^8}y^6 + \frac{5666}{3^9}y^7 + O(y^8) \nonumber \\
& \simeq & 1 + 0.333y - 0.111y^2 + 0.728y^3 + 0.181y^4 \nonumber \\
 & & - 0.044y^5 - 0.093y^6 + 0.288y^7 + O(y^8)
\label{wbarseriessqw2}
\eeqs
These may be compared with the first few terms of the series for the infinite 
lattice \cite{series}
\beq
\overline W(sq,q) = 1 + y^3 + y^7 + O(y^8)
\label{wbarsq}
\eeq
We see that as the width of the strip increases, the coefficients in the
small-$y$ series expansion of the corresponding $\overline W$ function 
approach those of
the infinite lattice, up to terms of order $O(y^p)$, where $p \sim L_y$. 
In particular, the coefficients of the terms $y^k$, $k=1,2,4,5,6$ are smaller
than those of the $y^3$ and $y^7$ terms. Evidently, as $L_y \to \infty$, these 
smaller coefficients eventually decrease to zero, while the coefficients of 
the $y^3$ and $y^7$ terms increase to their infinite-lattice values of unity.
For progressively larger strips, similar behavior occurs for the coefficients
of higher terms in the small-$y$ series.  In this way
one can see from small-$y$ series expansions of $\overline W$ functions on 
infinitely long strips how the infinite-lattice series arise. 

   It is also of interest to see how the values of the $W(\{G_{sq(L_y)}\},q)$ 
functions for the strips of various widths $L_y$ compare with those of the 
infinite square lattice, $W(\{G_{sq(L_y=\infty)}\},q) \equiv 
W(sq,q)$ for various values of $q$.  For example, for $q=3$, we have 
$W(\{G_{sq(1)}\},3)=2$, $W(\{G_{sq(2)}\},3)=\sqrt{3}=1.73205...$, 
$W(\{G_{sq(3)}\},3)=[(1/2)(5+\sqrt{17})]^{1/3}=1.65846...$, and 
$W(\{G_{sq(4)}\},3)=1.624945...$
These decrease monotonically toward the infinite-lattice value \cite{lieb}
$W(\{G_{sq(\infty)}\},3) \equiv W(sq,3) = (4/3)^{3/2}= 1.53960...$.  
Defining the ratio
\beq
R_W(\{G_{sq(L_y)}\},q) = \frac{W(\{G_{sq(L_y)}\},q)}{W(\{G_{sq(\infty)}\},q)}
\label{rw}
\eeq
we have $R_W(\{G_{sq(1)}\},3)=1.2990...$, $R_W(\{G_{sq(2)}\},3)=1.12500...$, 
$R_W(\{G_{sq(3)}\},3)=1.07720...$, and $R_W(\{G_{sq(4)}\},3)=1.05543...$. Thus 
the infinitely long strip graph of the square lattice with width $L_y=4$ 
already yields a $W$ function which is, for $q=3$, rather close to the value 
for the infinite square lattice.  We have verified that 
this is also true for higher values of $q$, using our Monte Carlo measurements
of $W(sq,q)$ in (Table 1 of) Ref. \cite{w}.  We have also verified that strips
graphs of other lattices show similar behavior.  We have found that, for a
fixed $q \ge q_c(\Lambda)$, $R_W(\{G_{s(L_y)}\},q)$ decreases toward
1 as the strip width increases.  Here,
following our discussion in Refs. \cite{w,w3,wn}, $q_c(\Lambda)$ denotes the 
maximal (finite) real value of $q$ where $W(\Lambda,q)$ is nonanalytic; in the
above works, we obtained the values $q_c(sq)=3$, $q_c(t)=4$, and 
$q_c(hc)=(3+\sqrt{5})/2=2.618..$ for the square, triangular, and honeycomb
lattices, respectively (see further discussion below on boundary conditions).

\vspace{3mm}

\begin{figure}
\centering
\leavevmode
\epsfxsize=3.5in
\begin{center}
\leavevmode
\epsfxsize=3.5in
\epsffile{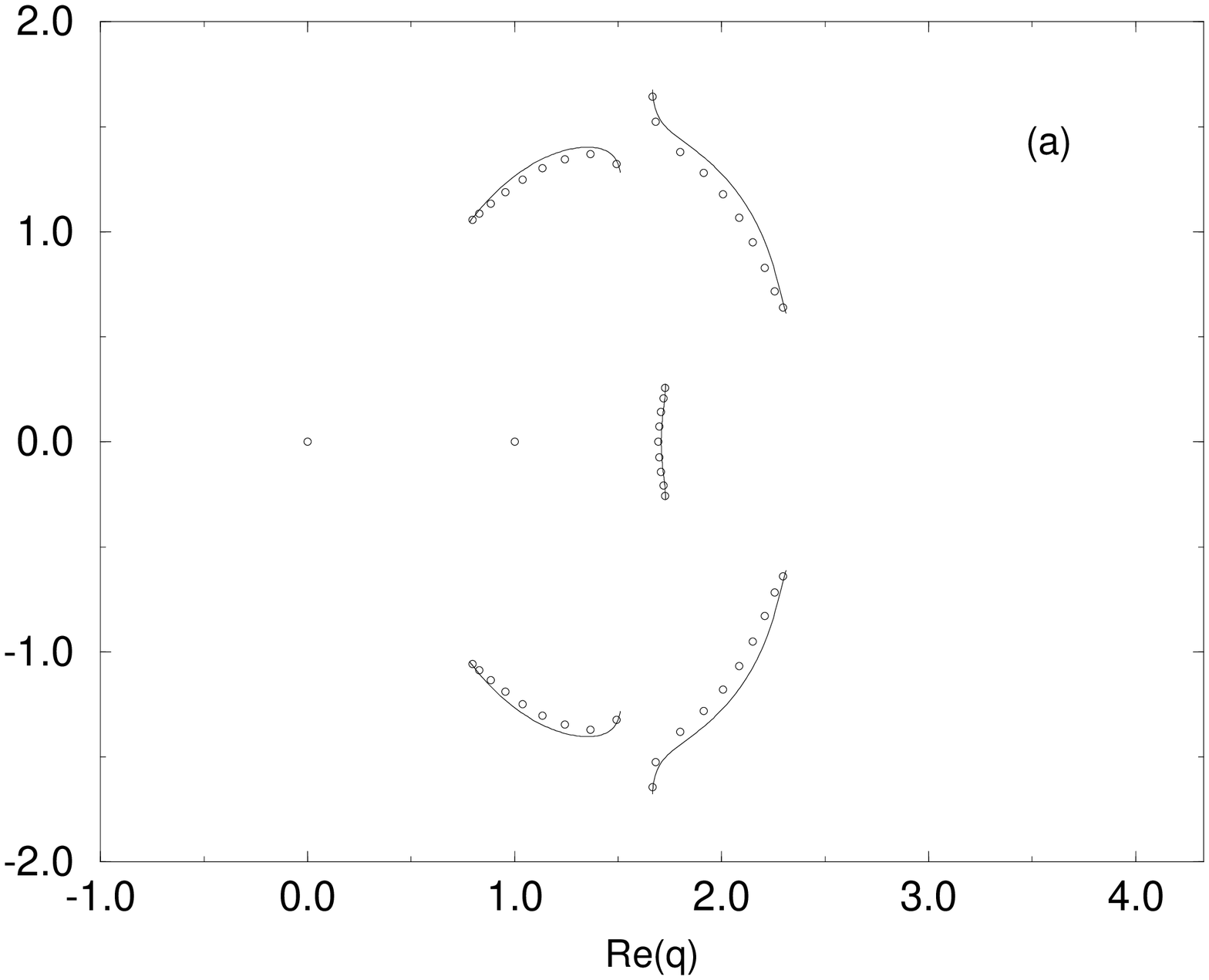}
\end{center}
\vspace{-4cm}
\begin{center}
\leavevmode
\epsfxsize=3.5in
\epsffile{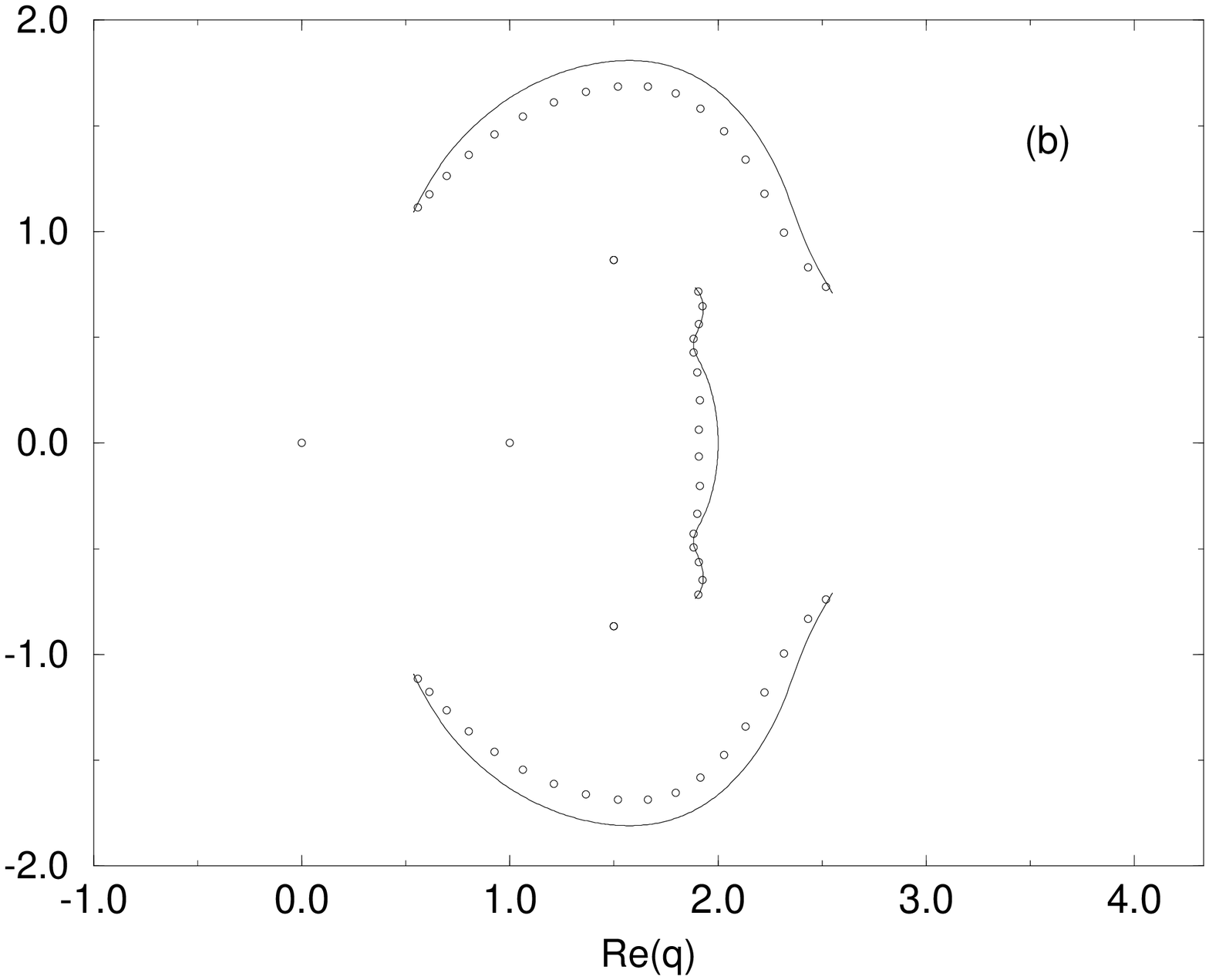}
\end{center}
\caption{\footnotesize{
Analytic structure of the function $W(\{G_{sq_d(L_d)}\},q)$ for 
$L_d=$ (a) 3, (b) 4, 
where $\{G_{sq_d(L_d)}\}$ denotes an infinitely long strip along a diagonal
direction of the square lattice, for the case where the repeating subgraph 
$H$ has diagonal width $d=L_d$.  Zeros of the chromatic polynomial
$P((G_{sq_d(L_d)})_m,q)$ for (a) $L_d=3$, $m=9$ (so $n=49$) and (b) 
$L_d=4$, $m=8$ (so $n=56$) are shown.}}
\label{figdiag}
\end{figure}
   
   In Figs. \ref{figdiag}(a) and \ref{figdiag}(b) we show our exact 
calculations of ${\cal B}$ for infinitely long strips of the square lattice 
along the diagonal direction.  Finite strips of these two respective types were
shown in Fig.\ \ref{figstrip}(c,d).  Comparing the two different-width strips,
one sees that the two pairs of complex-conjugate arcs in the narrower strip 
merge together to form a single such pair on the wider strip, while the 
self-conjugate arc elongates.  Furthermore, from a comparison of these two 
figures with the previous two, one sees several similarities but
also observes that even for a given lattice, the details of ${\cal B}$ depend,
in general, on the orientation of the strip with respect to the principal
lattice directions.  

\vspace{3mm} 

\begin{figure}
\centering
\leavevmode
\epsfxsize=3.5in
\begin{center}
\leavevmode
\epsfxsize=3.5in
\epsffile{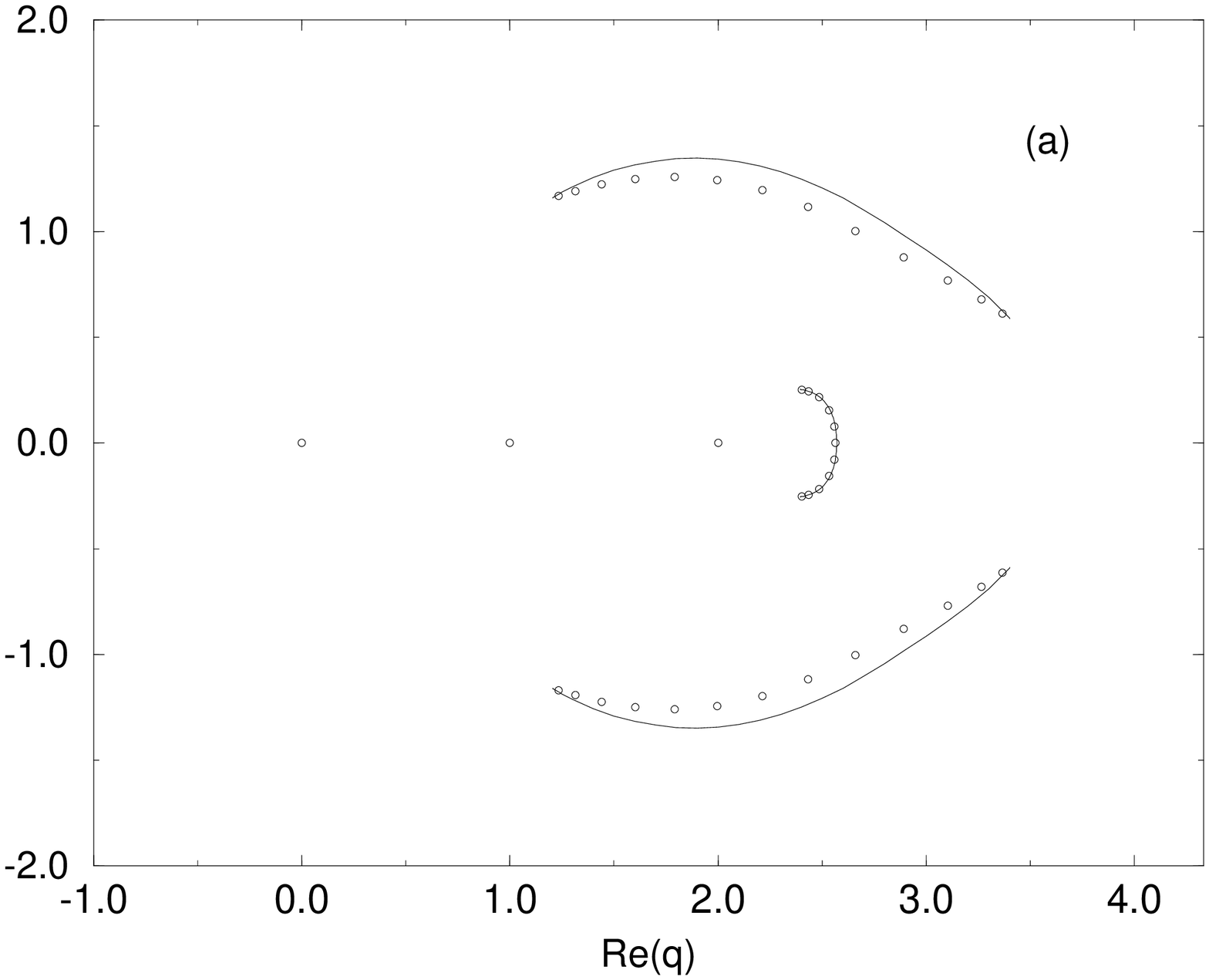}
\end{center}
\vspace{-4cm}
\begin{center}
\leavevmode
\epsfxsize=3.5in
\epsffile{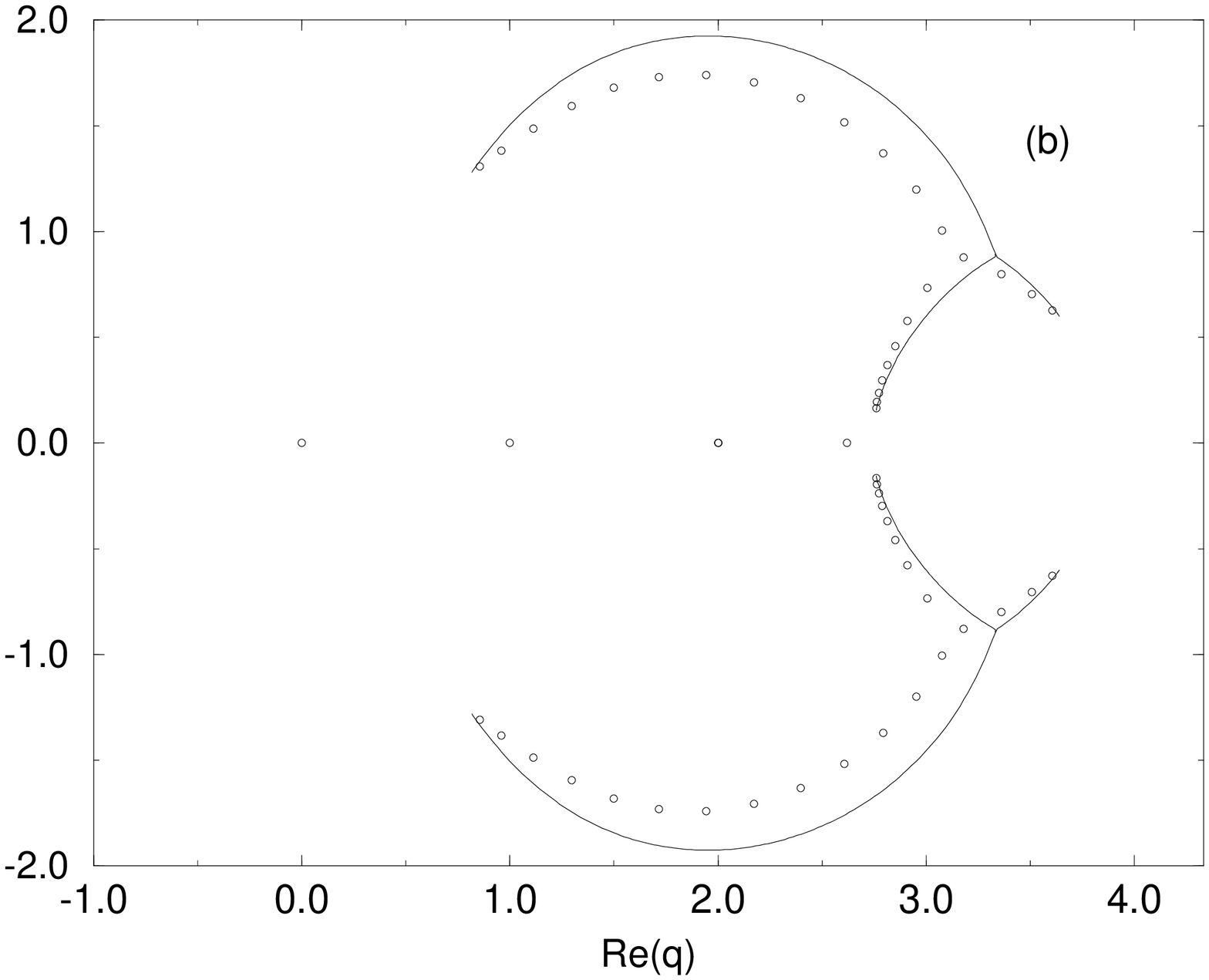}
\end{center}
\caption{\footnotesize{
Analytic structure of the function $W(\{G_{t(L_y)}\},q)$ for
$L_y=$ (a) 3, (b) 4, where $\{G_{t(L_y)}\}$ denotes the 
$(L_x=\infty \times L_y$ strip of the triangular lattice. 
For comparison, the zeros of the chromatic polynomials 
$P((G_{t(L_y)})_m,q)$ for (a) $L_y=3$, $m=12$ (hence, $n=42$) and 
(b) $L_y=4$, $m=12$ (hence $n=56$) are shown.}}
\label{figtri}
\end{figure}

   In Figs. \ref{figtri}(a) and \ref{figtri}(b) we show similar results for
strips of width $L_y=3$ and $L_y=4$ of the triangular lattice.  One sees a 
significant change in ${\cal B}$ as the width increases from 3 to 4; 
the self-conjugate arc crossing the real axis at $q=2.56984...$ for $L_y=3$ 
disappears, while what were previously 
the right-hand ends of the complex-conjugate arcs become multiple points of
index 3, each sprouting two additional arcs pointing in toward the real
axis.  The patterns of chromatic zeros for the long finite strips also 
change in the same manner, so that, as expected, for each width, these zeros
lie close to the respective asymptotic loci ${\cal B}$. Our explicit result for
$W$ for the $L_y=3$ strip is 
\beqs
W(\{G_{t(3)}\},q) & = & 2^{-1/3}\biggl [ q^3-7q^2+18q-17 + 
\nonumber \\
& & \Bigl [q^6-14q^5+81q^4-250q^3+442q^2-436q+193 \Bigr ]^{1/2} \biggr ]^{1/3}
\label{wtw2}
\eeqs
for real $q$ to the right of the crossing at $q=2.56984$ noted above, with
appropriate analytic continuation to the rest of the $q$ plane. 

\vspace{2mm}

\begin{figure}
\centering
\leavevmode
\epsfxsize=4.0in
\epsffile{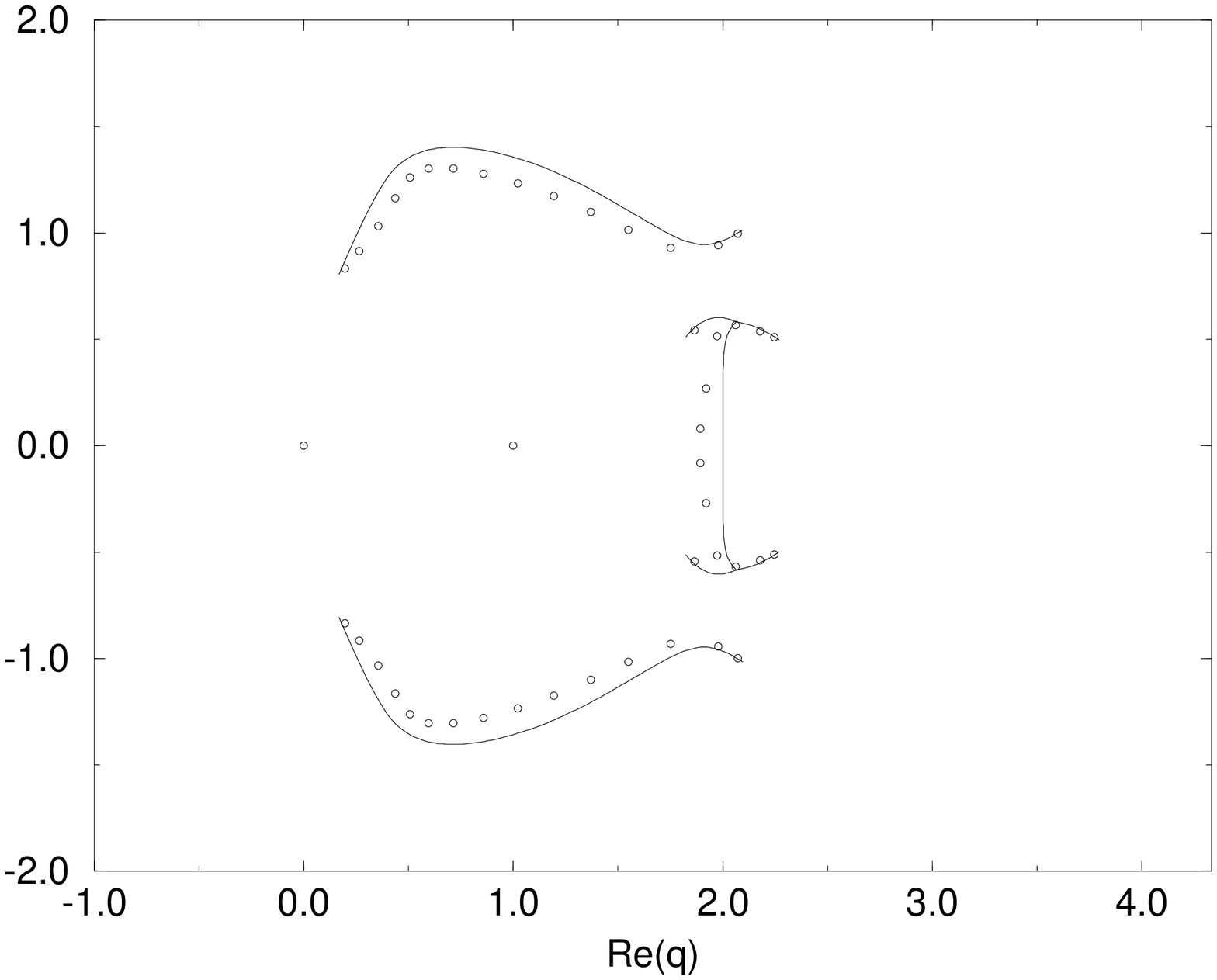}
\caption{\footnotesize{
Analytic structure of the function $W(\{G_{hc(3)}\},q)$, where 
$\{G_{hc(3)}\}$ the $L_x \times L_y = \infty \times 3$ strip of the honeycomb
(brick) lattice. Zeros of the chromatic polynomial $P((G_{hc(3)})_6,q)$ 
($n=46$) are shown.}}
\label{fighc}
\end{figure}

    In Fig.\ \ref{fighc} we show our calculations of ${\cal B}$ for a $L_x
\times L_y = \infty \times 3$ strip of the honeycomb (brick) lattice. 
As was true with the square and triangular lattices, the arcs help one 
understand the pattern of chromatic zeros calculated for an $8 \times 8$ 
lattice with the same free boundary conditions in Ref. \cite{baxter}; the 
positions of these arcs are close to the positions of the chromatic zeros, 
both for long finite strips and the $8 \times 8$ patch.  Furthermore, the
cusp-like structures in the latter case occur approximately where (i) our 
exact arcs have multiple points and (ii) there are gaps between the arcs.  

\begin{figure}
\centering
\leavevmode
\epsfxsize=4.0in
\epsffile{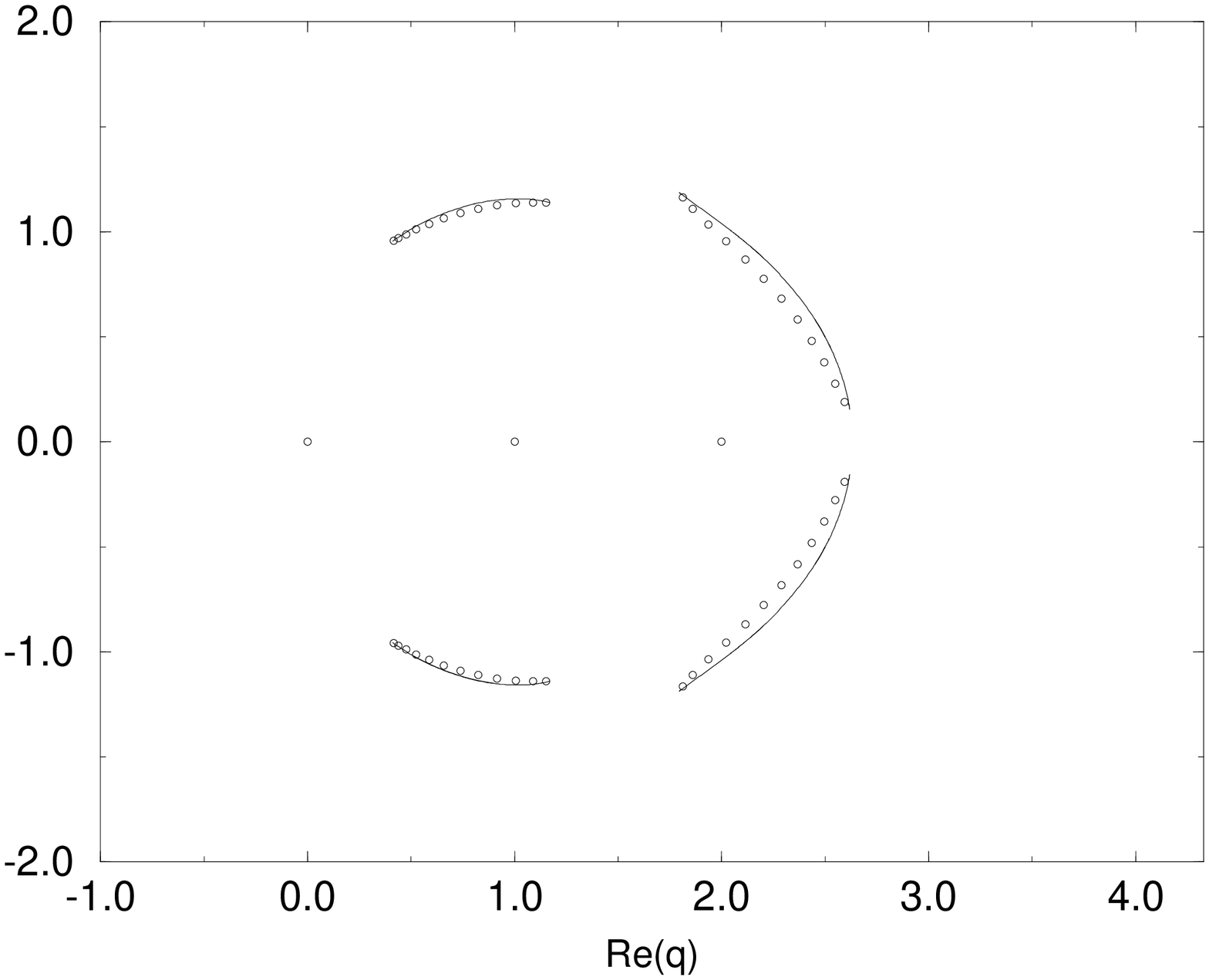}
\caption{\footnotesize{
Analytic structure of the function $W(\{G_{kag(2)}\},q)$, where 
$\{G_{kag(2)}\}$ denotes an infinitely long strip of the kagom\'e lattice, 
a finite-length section of which is shown in Fig.\ \ref{figstrip}(f). 
Zeros of the chromatic polynomial $P((G_{kag(2)})_9,q)$ of a 
finite strip with $m=9$ (hence $n=51$) are shown.}}
\label{fig3636}
\end{figure}

   Using our calculations of the generating functions, we have also
determined the exact $W$ functions and their associated continuous loci of
nonanalyticities ${\cal B}$ for infinitely long strips of three 
heteropolygonal lattices, {\it viz.}, kagom\'e 
($= (3 \cdot 6 \cdot 3 \cdot 6)$), 
$(3 \cdot 12^2)$, and $(4 \cdot 8^2)$, in the standard mathematical notation of
eq.\ (\ref{lambda}).  As an example, we consider the kagom\'e lattice.  In 
Fig.\ \ref{fig3636} we show our exact determination of ${\cal B}$ for an
infinite strip of this type, a finite section of which was shown in Fig.\ 
\ref{figstrip}(f).  There is a particular feature of this calculation that is 
worth noting: for strip graphs which have 
${\cal D}$ functions that are quadratic in $x$, the $\lambda_{s,k}(q)$'s given 
by eq. (\ref{lambdak2}) do not, in general, factorize.  However, for the
strip of the kagom\'e lattice considered here, the $\lambda_{s,k}(q)$'s do 
factorize.  This happens
because $b_{kag(2),1}$ and the square root of the discriminant both have a
common factor, $(q-2)$.  Related to this, not all of the zeros of the
discriminant are branch point singularities of the square root of this 
discriminant; the zero at $q=2$ is of even 
order (a double zero), and hence is not associated with a branch point of the
square root of the discriminant (although it is a branch point singularity of
the full $W$ function; see below).  We calculate 
\beqs
W(\{G_{kag(2)}\},q) & = & 2^{-1/5}(q-2)^{1/5}\biggl [ q^4-6q^3+14q^2-16q+10 
\nonumber \\
& & + \Bigl [q^8-12q^7+64q^6-200q^5+404q^4-548q^3+500q^2-292q+92
  \Bigr ]^{1/2} \biggr ]^{1/5}
\label{wkagw2}
\eeqs
As before, one can obtain this for positive real $q$ and analytically continue
the expression to the full complex $q$ plane, except for the locus ${\cal B}$
where $W(\{G_{kag(2)}\},q)$ is nonanalytic.  Note that although the point $q=2$
is a branch point singularity of $W(\{G_{kag(2)}\},q)$, it is an isolated
singularity, and does not lie on the continuous locus of nonanalytic points
${\cal B}$.

\vspace{4mm}

\begin{figure}
\centering
\leavevmode
\epsfxsize=4.0in
\epsffile{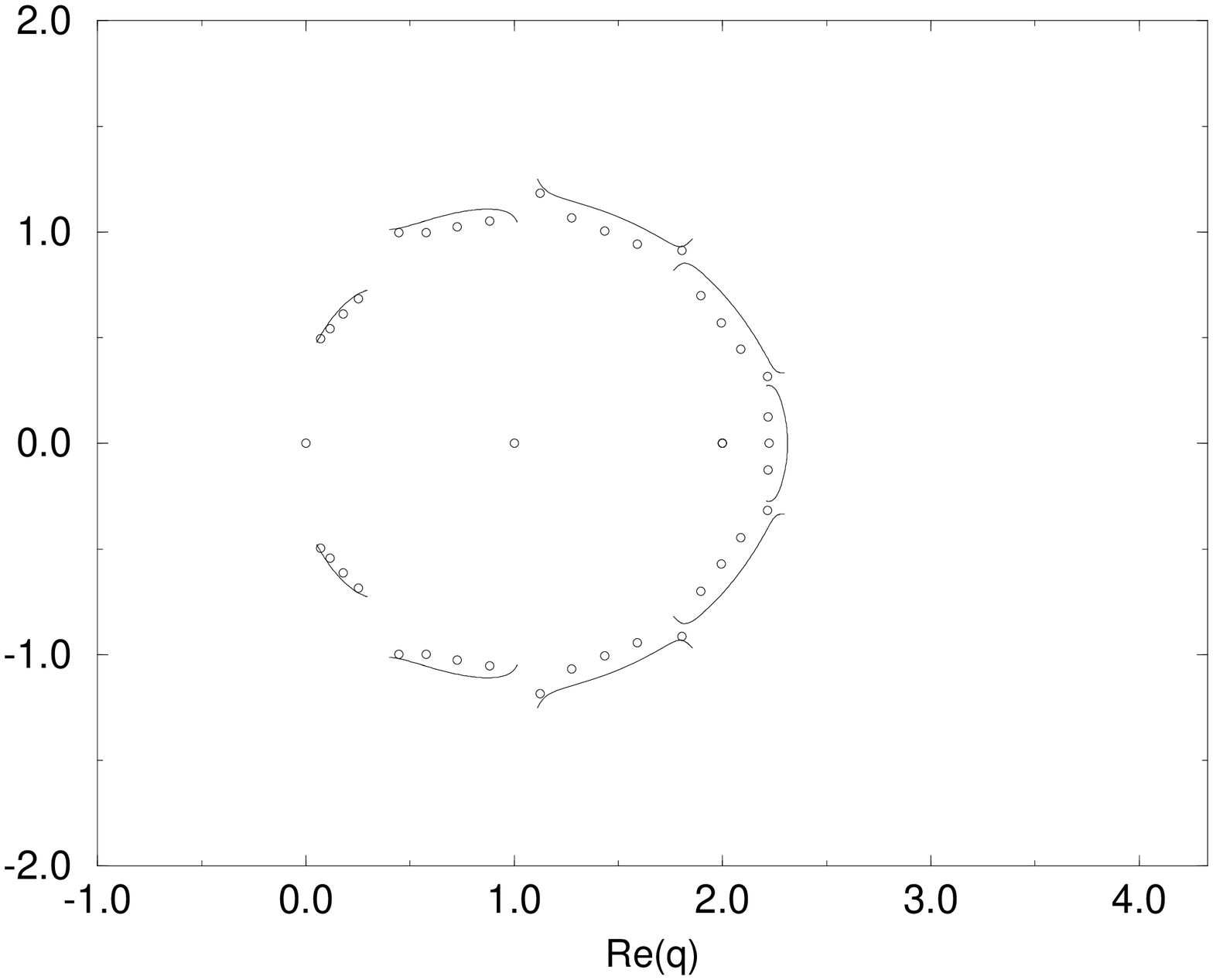}
\caption{\footnotesize{
Analytic structure of the function $W(\{G_{(3 \cdot 12^2)}\},q)$,
where $\{G_{3 \cdot 12^2)}\}$ denotes an infinitely long strip of the 
$(3 \cdot 12^2)$ lattice oriented as discussed in the text, and of width one
layer of 12-gons. For comparison, the zeros of the chromatic polynomial
$P(G_{(3 \cdot 12^2)})_3,q)$ of a finite strip with $m=3$ (hence $n=42$ 
vertices), are shown.}}
\label{fig312}
\end{figure}

\begin{figure}
\centering
\leavevmode
\epsfxsize=4.0in
\epsffile{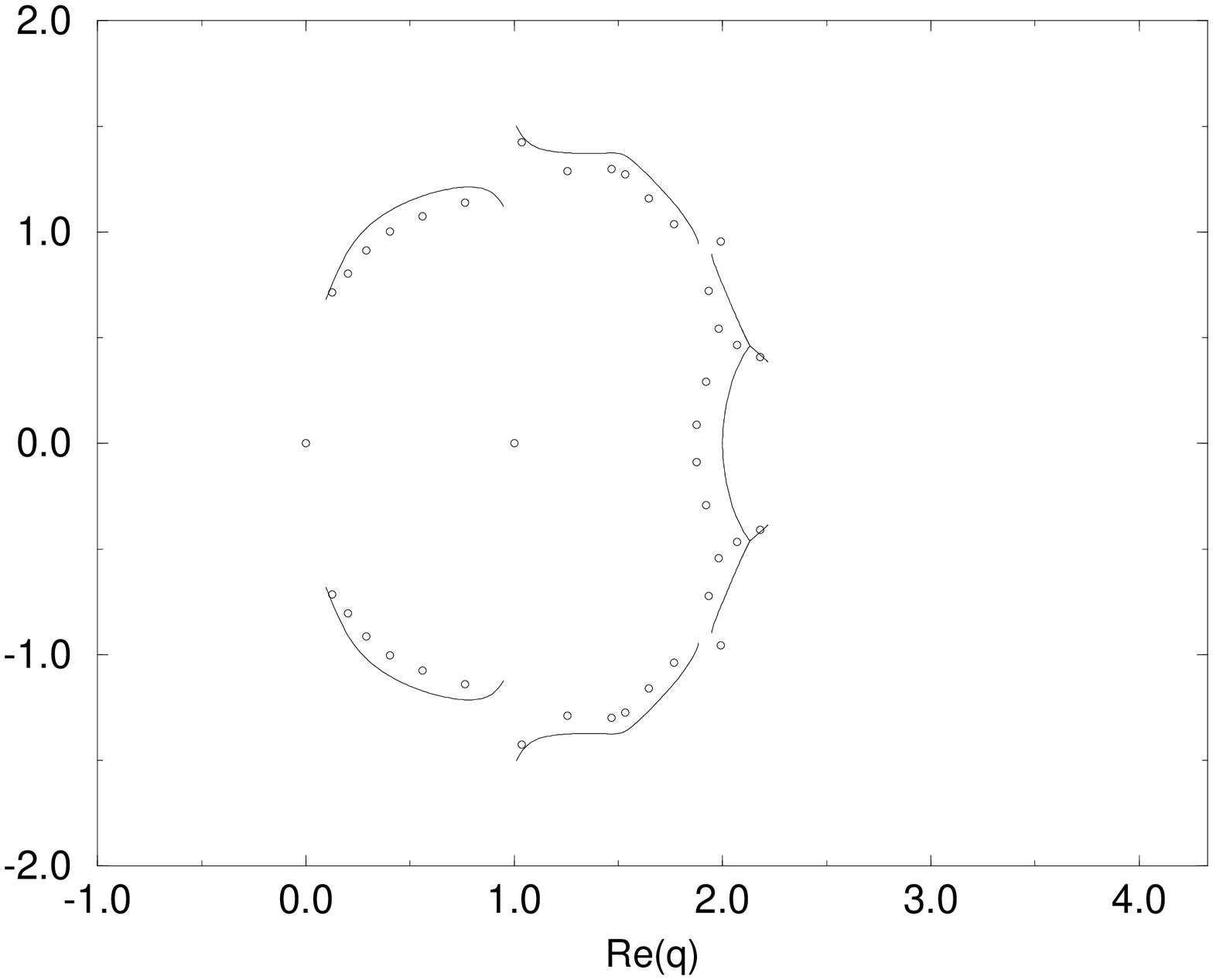}
\caption{\footnotesize{Analytic structure of the function 
$W(\{G_{(4 \cdot 8^2)(2)}\},q)$, where 
$\{G_{(4 \cdot 8^2)(w)}\}$ denotes an infinitely long strip of the 
$(4 \cdot 8^2)$ lattice of width $w$; a finite-length section is 
shown in Fig.\ \ref{figstrip}(h). 
For comparison, the zeros of the chromatic polynomial 
$P((G_{(4 \cdot 8^2)(3)})_2,q)$ of a finite strip (with $n=40$) is shown.}}
\label{fig488}
\end{figure}

   We present our results for the generating functions of the other
heteropolygonal Archimedean lattices in Appendix 2 and the 
resultant determinations of ${\cal B}$ in the figures of Appendix 3.  
The chromatic zeros for long finite strips of each of these types are also 
shown in these figures.  As with homopolygonal lattices,
the arcs for the $w=1$ $(3 \cdot 12^2)$ and $w=2$ $(4 \cdot 8^2)$ infinite 
strips show close similarity in position not only with the chromatic zeros 
calculated for finite strips of the same type and width but
also with our previous calculations of chromatic zeros for reasonably large
patches of the respective lattices with $L_x \simeq L_y$ in Ref. \cite{wn}.  
For the $(3 \cdot 12^2)$ lattice, the large number (nine) of closely spaced 
arcs in Fig.\ \ref{fig312} correlates nicely with the rather smooth and round 
distribution of chromatic zeros that we presented in Fig.\ 8 of Ref. \cite{wn}.
For the $(4 \cdot 8^2)$ lattice, the correspondence with the chromatic zeros 
of Fig.\ 9 in Ref. \cite{wn} is close for the $w=2$ strip, shown in
Fig.\ \ref{fig488}; here, as before, one sees a faithful correspondence
between cusp-like structures in the chromatic zeros on the 2D path and the
multiple points and gaps between arcs on ${\cal B}$ for the strip.

\vspace{3mm}

   There are a number of interesting features exhibited by our results. 
A general feature is that for all of these strip graphs of the form 
$G_s = (\prod_{\ell=1}^m H)I$ as in eq. (\ref{gs}), the respective 
loci ${\cal B}$ form arcs which do not separate the complex $q$ plane into 
different, disconnected regions.  This finding may be compared with our 
exact results in Refs. \cite{w,wa}.  There we showed 
that the nature of the locus ${\cal B}$ depends strongly on the topology of a 
graph, (boundary conditions for a lattice graph).  
In the simplest case of an $n$-vertex tree graph $T_n$, in the 
limit $n \to \infty$, $W(\{T\},q)$ is analytic in the entire finite $q$ 
plane ({\it i.e.}, ${\cal B}=\emptyset$, the null set), whereas if
one ties together the ends to make an $n$-vertex circuit graph $C_n$, then in
the limit $n \to \infty$, $W(\{C\},q)$ is nonanalytic on the boundary 
${\cal B}$ formed by the circle $|q-1|=1$, which divides the $q$ plane into two
regions \cite{w}.  Similarly, the $n \to \infty$ limit of an $n$-vertex strip 
graph (chain) composed of $m$ $p$-sided polygons, denoted $(Ch)_{p,n}$ in 
Ref. \cite{wa} (with $n=(p-2)m+2$) yields a $W(\{(Ch)_p\},q)$ function with
${\cal B}=\emptyset$, whereas if one ties together the ends to form a 
circular ladder graph, the resultant function $W(\{L\},q)$ is nonanalytic 
on a continuous locus ${\cal B}$ which forms the
boundaries of four different regions in the $q$ plane (as shown in Fig.\ 3 of 
Ref. \cite{w}).  In all of these cases, ${\cal B}$ was either (i) the null 
set $\emptyset$, for the $n \to \infty$ limit of $n$-vertex 
graphs such as the complete graph $K_n$, tree graph $T_n$, and 
strip (chain) graph composed of $p$-gons, $(Ch)_{p,n}$, or (ii) a set of
boundaries that separated the complex $q$ plane into two or more different
regions, within which $W(\{G\},q)$ was analytic, such as for the circuit, 
$C_n$, periodic and twisted ladder graphs, and others like the wheel graph
and its generalization to the $p$-wheel graph \cite{w,wa}\footnote{  
The $n$-vertex wheel graph $(Wh)_n$ is the graph formed by 
adjoining a point ({\it i.e.}, connecting it with bonds) to all vertices of an 
$(n-1)$-vertex circuit graph $C_{n-1}$.  This may be denoted 
$(Wh)_n = K_1 \times C_{n-1}$, where $K_\ell$ is the complete graph on $\ell$
vertices.  The generalized $p$-wheel graph is given by 
$(Wh)^{(p)}_n = K_p \times C_{n-p}$ with $(Wh)^{(1)}_n \equiv (Wh)_n$ 
\cite{w,wa}.}.  Thus, the strip graphs studied here yield $W(\{G_s\},q)$ 
functions with analytic structure 
intermediate between type (i) and (ii) found earlier: ${\cal B}$ is nonempty
but does not divide the $q$ plane into separate regions of analyticity. 
Moreover, different strips exhibit a wide variety in the number of arcs.  
As we shall discuss in Ref. \cite{strip2}, we have found that for strip 
graphs of the form $J(\prod_{\ell=1}^m H)I$, ${\cal B}$ contains arcs
again, but these can actually enclose regions; this happens if 
the degeneracy equation (\ref{degeneracy}) involves at least one 
$\lambda_{s,k}$ which is a
polynomial in $q$.  In contrast, for all of the strip graphs of the form 
$(\prod_{\ell=1}^m H)I$ and having ${\cal B} \ne \emptyset$ considered in the 
present paper, the maximal $\lambda_{s,k}$'s that occur in the degeneracy 
equation (\ref{degeneracy}) are algebraic, but not simply polynomial, 
functions of $q$, {\it i.e.}, they involve branch point singularities.  

   The origin of this arc structure of ${\cal B}$ for the strip graphs studied
here can be seen explicitly by examining the calculation of this locus in the 
simple case where $k_{max}=2$, {\it i.e.}, where the denominator 
${\cal D}(G_s,q,x)$ of the generating function 
$\Gamma(G_s,q,x)$ is quadratic as a polynomial in $x$. Here, aside from the 
trivial case where for some $q$, $b_{s,1}(q)=b_{s,2}(q)=0$ \footnote{
The restriction to nonvanishing $b_{s,k}$ is relevant for 
the ${\cal D}$ functions for the kagom\'e ({\it i.e.}, 
$(3 \cdot 6 \cdot 3 \cdot 6)$) and $(3 \cdot 12^2)$ strips of width $L_y=2$; 
since both of these strips contain triangles, the chromatic polynomials
vanish for all $m \ge 1$ if $q=2$.  This is reflected in the property that for
$q=2$, the coefficient functions $a_{s,1}=b_{s,1}=b_{s,2}=0$, and the 
respective generating functions become $x$-independent.  Hence, as is evident 
from Figs. \ref{fig3636} and \ref{fig312}, the point $q=2$ is not on the
nonanalytic locus ${\cal B}$.  Related to this, as we have indicated in Table
1, in these two cases, although the respective degrees of $R_s$ are 10 and 20,
the number $N_{b.p.}$ of branch points of $\sqrt{R_s}$ are less by 2, 
{\it viz.}, 8 and 18. This in contrast to the other strips listed in Table 1, 
where $deg(R_s)=N_{b.p.}$.} the continuous locus of points ${\cal B}$ where 
$W(\{ G_s\},q)$ is nonanalytic is 
determined by the special case of eq.\ (\ref{degeneracy}) with 
$\lambda_{s,k}(q)$, $k=1,2$ given by eq.\ (\ref{lambdak2}). 
The arcs comprising ${\cal B}$ extend between, and end on, zeros of the 
discriminant.  It follows that $N_{arc} = N_{b.p.}/2$, where $N_{arc}$ and
$N_{b.p.}$ denote the number of arcs and branch points, respectively.  In 
Table 1 we have listed these for the strip graphs with $k_{max}=2$ considered
here.  An analogous connection holds 
between the arc endpoints and branch points in the maximal-magnitude
$\lambda_{s,k}$'s appearing in the degeneracy equation (\ref{degeneracy}) 
for the case $k_{max}=3$, and one can see that it is a general feature of the
degeneracy equations for $\lambda_{s,k}$'s of maximal magnitude.  
Because ${\cal B}$ is invariant under complex conjugation, it follows that 
the arcs must occur as complex-conjugate pairs together with possible
self-conjugate arcs.  Clearly if $N_{arc}$ is odd, then 
there must exist at least one such self-conjugate arc.  One can see by scanning
Table 1 and the associated figures various examples illustrating both the
presence and absence of self-conjugate arcs.

   The fact that ${\cal B}$ consists of arcs ending on branch points of the
maximal-magnitude $\lambda_{s,k}$ in the complex $q$ plane is reminiscent of 
the appearance of arcs (and line segments) of nonanalyticities of the free 
energy, ending on branch points of maximal-magnitude eigenvalues of the 
transfer matrix, in exact calculations of complex-temperature phase diagrams 
of 1D systems (e.g., \cite{z6,is1d,1dnnn}) and finite-width strips 
\cite{martin}.  Our earlier work on complex-temperature phase diagrams 
\cite{z6,is1d,1dnnn}
helps to elucidate the conditions under which arcs or line segments do or do
not enclose distinct regions in the respective complex plane (here, the $q$
plane, there the complex plane in an appropriate Boltzmann weight variable). 
For example, in the 1D nearest-neighbor Ising model with spin $s=1/2$, the 
eigenvalues of the transfer matrix are given by 
$\lambda_{1/2,\pm}=u^{-1/4}(1 \pm \sqrt{u})$, and hence ${\cal B}$ consists 
of the semi-infinite line segment $\infty \le u \le 0$, where $u=e^{-4K}$ with 
$K=\beta J$, $\beta=1/(k_BT)$, and $T$ and $J$ are the temperature and
spin-spin coupling.  This locus of points ends on the branch point $u=0$ and
does not separate the complex $u$ plane into different regions. 
In contrast, for spin $s=1$, the eigenvalues of the transfer matrix are 
$\lambda_{1,1}=u_1^{-1}-u_1$ and
\beq
\lambda_{1,\pm}=(1/2)\Bigl [ u_1^{-1}+1+u_1 \pm
(u_1^{-2}-2u_1^{-1}+11-2u_1+u_1^2)^{1/2} \Bigr ]
\label{lambda123}
\eeq
where $u_1=e^{-K}$; here the curves comprising ${\cal B}$, which
are the solutions to the degeneracy equation of maximal eigenvalues of the
transfer matrix $|\lambda_{1,1}| =|\lambda_{1,+}|$, do separate the
complex $u_1$ plane into (four) different regions, as shown in Fig.\ 1(a) of
Ref. \cite{is1d}.   These curves end on the singular point at $u_1=0$ and
extend to complex infinity.  Note that no arcs end on the finite branch 
points of the square root in eq.\ (\ref{lambda123}) because neither 
$\lambda_{1,+}$ nor $\lambda_{1,-}$ is a dominant eigenvalue at any of these
points.  Related to this, when one crosses one of the curves, one does not
switch between two $\lambda$'s that are related only by changing the sign of
the square root, in contrast, e.g., to the situation described by eq.\ 
(\ref{degeneracy}) for strip graphs considered here.  One also encounters
arcs which close in such a manner as to separate the complex-temperature plane
into different regions which are compact, rather than extending to infinity.
This occurs, for example, in the case of the 1D $\Z_6$ clock model \cite{z6}
and the 1D spin 1/2 Ising model with nearest- and next-nearest-neighbor
spin-spin couplings $J_{NN}$ and $J_{NNN}$ for certain values of the ratio 
$J_{NNN}/J_{NN}$ \cite{1dnnn}. 

   Another salient feature of our results is that although in some cases there
is often a self-conjugate arc which crosses the real $q$ axis farthest to the 
right, such an arc does not occur farthest to the left.  This is also reflected
in the reduced density (along arcs) of chromatic zeros near to $q=0$ 
(aside from the zero precisely at $q=0$, which is always present for any 
chromatic polynomial).  

     Our current results are relevant to the question of the 
conditions under which components of ${\cal B}$ are
compact or, in contrast, extend infinitely far from the origin of the $q$
plane, yielding a reduced function $\overline W(\{G\},q)$ which is 
nonanalytic at $1/q=0$.   A family of graphs with noncompact ${\cal B}$ was
studied in Refs. \cite{read91,w}, and in Ref. \cite{wa} infinitely many 
families of graphs were constructed with noncompact curves ${\cal B}$, leading
to the formulation of a general condition 
explaining when such noncompactness does or does not occur. 
Since for the $G_s$ strip graphs considered here the branch 
points of the maximum-magnitude $\lambda_{s,k}$'s, and hence the endpoints of 
the arcs, occur at finite values of $q$, these arcs never extend infinitely far
from the origin $q=0$.  This is in accord with the criterion for compactness of
${\cal B}$ presented in our earlier general study \cite{wa}. 

   A longstanding question in graph theory has concerned general features of
chromatic zeros for a general graph $G$.  Given the merging of a subset of 
these in the infinite-vertex limit to form the locus ${\cal B}$ of points 
where $W(\{G\},q)$ is nonanalytic, one is also led to ask about general
features of ${\cal B}$ for an arbitrary graph.  We have discussed several of
these above. Here we formulate a conjecture concerning the sign of the real
part of chromatic zeros and thus of points on ${\cal B}$.  Before doing this,
we recall that some time ago, it was conjectured \cite{farrell} that the
chromatic zeros of an arbitrary graph have non-negative real parts.  This
conjecture was subsequently shown to be false, and now many examples of graphs
are known with chromatic zeros having negative real parts.  
However, a striking feature of our exact 
calculations of chromatic zeros and nonanalytic loci ${\cal B}$ in Refs. 
\cite{w,wa,wn} for many families of graphs, as well as the calculations of 
chromatic zeros
for lattices with free, as opposed to cylindrical, boundary conditions, in
Ref. \cite{baxter}, is that they have $Re(q) \ge 0$. We are thus 
led to state the following conjectures.  We first give a definition: a regular
lattice graph has what we denote as a ``global circuit'' if and only if by
travelling along one of the lattice vectors one can arrive back at one's
starting point.  The definition of lattice vector is clear for a cartesian
$d$-dimensional lattice; for the triangular and honeycomb lattices, the lattice
vectors can be taken to lie along $0^\circ$, $60^\circ$, and $120^\circ$,
etc.  Thus, a patch of any of these lattices with free boundary conditions has
no global circuits, while a patch with periodic boundary conditions in at least
one direction has global circuits along this direction.  

\vspace{2mm}

\begin{flushleft}
Conjecture 1a
\end{flushleft}

If $q_0$ is a chromatic zero of a regular lattice graph $G$ with no global
circuits, then $Re(q_0) \ge 0$.  
Furthermore, the only chromatic zero with $Re(q_0)=0$ is $q_0=0$ itself. 

\begin{flushleft}
Conjecture 1b
\end{flushleft}

\vspace{3mm}

  Consider the $n \to \infty$ limit of a regular lattice graph $G$ with 
no global circuits.  The points $q \in {\cal B}$ have the property that 
$Re(q) \ge 0$.  Furthermore, the only point on
${\cal B}$ with $Re(q) = 0$ is the point $q=0$ itself. 

\vspace{3mm}

Note that our exact calculations in Refs. \cite{w,wa} show that a graph may 
have global circuits and still have chromatic zeros and, in the infinite-$n$
limit, a locus ${\cal B}$, with the properties that (i) $Re(q) \ge 0$ and (ii)
$Re(q) = 0 \ \Rightarrow \ q = 0$; this was true, in particular, of the circuit
graph, its generalization \cite{wa} to $K_p \times C_{n-p}$.  This is also 
true of the ladder graph (strip of the square lattice of width $L_y=2$ with 
ends tied together (in an untwisted or twisted manner), {\it i.e.}, with 
periodic boundary conditions in the longitudinal direction 
\cite{w}, both of which have global circuits along the longitudinal
direction. This shows that the condition of no global circuits is a 
sufficient, but not necessary, condition for 
properties (i) and (ii) above.  Although our previous (exact) calculations of
chromatic polynomials and $W$ functions in Refs. \cite{w,wa} showed that these
are affected by boundary conditions, it also showed that, in all the cases
considered, when one compared a given type of graph with different topology
({\it i.e.}, existence or nonexistence of global circuits) due
to different boundary conditions, there did exist a unique value $q_c$ such
that for real $q \ge q_c$, the $W$ functions calculated with the different
boundary conditions coincided.  For example, for the line graph with free
boundary conditions (and more generally the tree graph), in the infinite-vertex
limit, $W(\{T\},q)=q-1$, while for the circuit graph $C_n$ ({\it i.e.} the line
graph with periodic boundary conditions), in the same limit, $W(\{C\},q)=q-1$
for $|q-1| > 1$ and $|W(\{C\},q)|=1$ for $|q-1| < 1$ (see section IV of
Ref. \cite{w}). Thus, $W(\{T\},q)=W(\{C\},q)$ for real $q \ge q_c$, where 
$q_c=2$ in this case (more generally, they also coincide throughout the 
region $|q-1| \ge 1$ in the complex $q$ plane). 
Similarly, for the $n$-vertex ladder graphs $L_n$, in the 
$n \to \infty$ limit with free boundary conditions, $W(\{L\}_{fbc},q)=
(q^2-3q+3)^{1/2}$, while for (twisted or untwisted) periodic boundary
conditions, this expression again holds in the region denoted $R_1$ in
Ref. \cite{w}, defined in general as the maximal region which is analytically 
connected with the real interval $q \ge q_c$ for a given lattice\footnote{
In eqs. (4.22)-(4.24) of Ref. \cite{w}, the $W$ function was
defined per rung rather than per site; for the definition per site, as used
here, the right-hand sides of these equations are raised to the power $1/2$.}.
For the periodic ladder graphs, $q_c$ is again equal to 2 ($R_1$ also 
includes the interval $q < 0$ on the real axis), and there are three other
regions in the complex $q$ plane where $W$ takes on other analytic forms.
These exact results suggest the following generalization: for a given type of
strip graph $G_s$ of a regular lattice, 
for real $q \ge q_c$, where $q_c$ is the maximal (finite) real
value at which $W(\{G_s\},q)$ is nonanalytic (equivalently, ${\cal B}$
crosses the real axis) for some set of boundary conditions, the
functions $W(\{G_s\},q)$ calculated with the different sets of boundary
conditions coincide.  A further generalization is that these $W$ functions can
be stated as follows.  Consider the set of (in general different) loci 
${\cal B}$ calculated for the infinite-vertex limit of a regular lattice with
different boundary conditions.  Consider the maximal envelope of ${\cal B}$ for
this set.  Then the $W$ functions agree in the region outside this maximal
envelope, which region includes the area extending infinitely far away from the
origin of the $q$ plane, {\it i.e.}, the inverse image of the origin of the
$1/q$ plane. 
Note that this property is necessary for the existence of large-$q$ 
Taylor series expansions of $\overline W$ for regular lattices.  As discussed
in Refs. \cite{w,wa}, the existence of these large-$q$ series expansions for
$\overline W$ also depends on the property that in these cases ${\cal B}$ does
not extend infinitely far away from the origin in the $q$ plane. 

\section{Conclusions}

   In conclusion, in this paper we have calculated the chromatic polynomials 
$P(G_s,q)$ for strip graphs comprised of repeated subgraphs $H$ adjoined to 
an initial graph $I$ and, from these, in the limit of infinitely long strips,
the resultant $W(\{G_s\},q)$ functions. To do this, we have developed a
powerful method in which we compute a generating function $\Gamma(G_s,q,x)$ 
which yields these chromatic polynomials as coefficients in a series expansion
in $x$.  We have studied the analytic structure of $W(\{G_s\},q)$ for a number
of different types and widths of strip graphs.  The locus of points where
$W(\{G_s\},q)$ is nonanalytic consists of arcs which do not separate the 
complex $q$ plane into different regions.  From comparisons of arcs for
different-width strips of given types, we have shown how these arcs 
elongate and tend to move toward each other as the width of the infinitely 
long strips is increased.  We have calculated chromatic zeros
for long finite-length strips and have shown that, aside from the generally
occuring real zeros at $q=0,1$ and, for strips containing triangles, $q=2$,
these lie close to the positions of the arcs.  As we have discussed, these
calculations of the ground state degeneracy $W(\Lambda,q)$ of the $q$-state
Potts antiferromagnet on these infinitely long strips of various widths helps 
one to understand the behavior of this function on infinite 2D lattices 
$\Lambda$. 
In turn, this adds to our understanding of ground state degeneracy and
entropy, a phenomenon which is exhibited in the physical world by such
compounds as (H$_2$O) ice. 

\vspace{2mm} 

\begin{center}
{\bf Acknowledgments} 
\end{center}

This research was supported in part by the NSF grant PHY-93-09888.

\vspace{6mm}

\section{Appendix 1}

In this Appendix we present two alternate ways to calculate the generating
function $\Gamma(G_s,q,x)$ that are complementary to the method given in
section IIB. 

\subsection{Calculation of the Generating Function by Matching of Terms}

   Here, as before, we solve for $\Gamma(G_s,q,x)$ simply by matching 
the Taylor series expansion of $\Gamma(G_s,q,x)$ to
explicit calculations of the chromatic polynomials for the first few values of
$m$.  Since there are $N_p = j_{max}+k_{max}+1$
polynomials in (\ref{gammagen})-(\ref{d}), we can determine these by 
(i) calculating the chromatic polynomials $P((G_s)_m,q)$ for the various 
strips of type $G_s$ with $m$ ranging from 0 to $N_p$, 
(ii) performing the Taylor series expansion of $\Gamma(\{G_s\},q,x)$ about 
$x=0$ to order $N_p-1$, (iii) equating the successive terms in the Taylor 
series expansion to the respective chromatic polynomials, using 
(\ref{gamma}), and 
(iv) solving for the $a_j$, $j=0..,j_{max}$ and $b_k$, $k=1..,k_{max}$.
For example, for many strip graphs of width $w=L_y-1=2$, we find that 
$j_{max}=1$, $k_{max}=2$ \footnote{
The honeycomb strip graph is an exception; for $w=2$, the
resultant generating function has $j_{max}=2$, $k_{max}=3$.}, 
so that $\Gamma(G_s,q,x)$ has the form 
(\ref{gammamin}) given above.  Evaluating eq.\ (\ref{gamma}), one has 
\beq
a_{s,0} = P((G_s)_0,q)
\label{term0}
\eeq
(where $P((G_s)_0,q) = P(I,q)$ from eq.\ (\ref{piq}))
\beq
a_{s,1} - a_{s,0}b_{s,1} = P((G_s)_1,q)
\label{term1}
\eeq
\beq
a_{s,0}b_{s,1}^2 - a_{s,0}b_{s,2} - a_{s,1}b_{s,1} = P((G_s)_2,q)
\label{term2}
\eeq
and 
\beq
2a_{s,0}b_{s,1}b_{s,2} - a_{s,1}b_{s,2} + a_{s,1}b_{s,1}^2 - 
a_{s,0} b_{s,1}^3 = P((G_s)_3,q)
\label{term3}
\eeq
which has the solution given by (\ref{term0}) together with 
\beq
a_{s,1} = \frac{2P((G_s)_0,q)P((G_s)_1,q)P((G_s)_2,q)-
P((G_s)_0,q)^2P((G_s)_3,q)-P((G_s)_1,q)^3}{Q}
\label{a1sol}
\eeq
\beq
b_{s,1} = \frac{P((G_s)_1,q)P((G_s)_2,q)-P((G_s)_0,q)P((G_s)_3,q)}{Q}
\label{b1sol}
\eeq
\beq
b_{s,2} = \frac{P((G_s)_1,q)P((G_s)_3,q)-P((G_s)_2,q)^2}{Q}
\label{b2sol}
\eeq
where
\beq
Q = P((G_s)_0,q)P((G_s)_2,q)-P((G_s)_1,q)^2
\label{q}
\eeq
The resulting $a_1$, $b_1$, and $b_2$ are polynomials, although this is not
manifest in eqs. (\ref{a1sol})-(\ref{b2sol}).

\subsection{Calculation of the Generating Function by Addition-Contraction
Theorem}

   Another alternate method of calculation is based on the 
addition-contraction theorem from graph theory.  This method is quite powerful
and allows us also to calculate chromatic polynomials and asymptotic limiting
functions $W$ for strip graphs with specific subgraphs at both ends, not just
one.  We first recall the statement of the addition-contraction theorem:
let $G$ be a graph, and let $v$ and $v'$ be two non-adjacent vertices
in $G$.  Form (i) the graph $G_{add.}$ in which one adds a bond connecting $v$
and $v'$, and (ii) the graph $G_{contr.}$ in which one identifies $v$ and
$v'$.  Then the chromatic polynomial for $G$ is equal to the sum of the
chromatic polynomial for the graphs $G_{add.}$ and $G_{contr.}$.  For our
proof, with no loss of generality, it is convenient to take the initial
subgraph $I$ in eq.\ (\ref{gs}) to be identical to the repeating subgraph
unit $H$; once having calculated the generating function for this case, it is
straightforward to obtain the generating function for the strip with an initial
subgraph $I$ attached to the right end.  By applying the addition-contraction 
theorem to the right-hand side of the initial $H$ subgraph in
$(G_s')_m$, we obtain a set of strip graphs $(G'_{s,j})_m$ with complete
subgraphs labeled by $j$ on the right-hand end. For example, let us consider a
strip $(G_s')_m = \prod_{\ell=1}^m H$ with $L_y=4$ vertices in the transverse
direction, and label the vertices on the right-hand end, in sequence, as 
$v_1,v_2,v_3,v_4$.  Now apply the addition-contraction theorem to the
pair $v_1,v_4$; this yields the equation
\beq
P((G_s')_m,q) = P((G'_{s,b(1,4)})_m,q) + P((G'_{s,v_1=v_4})_m,q)
\label{v1v4}
\eeq
where $(G'_{s,b(1,4)})_m$ denotes the strip graph with the right-hand end
modified by the addition of the bond connecting vertices $v_1$ and $v_4$, and
$(G'_{s,v_1=v_4})_m$ denotes the strip graph with the right-end modified by
identifying vertices $v_1$ and $v_4$.  Applying the theorem again to each of
these two strip graphs, one obtains the equation
\beqs
P((G'_s)_m,q) & = & P((G'_{s,K_4})_m,q) + P((G'_{s,K_3(v_1=v_4)})_m,q) +
P((G'_{s,K_3(v_1=v_3)})_m,q) \nonumber \\
& & + P((G'_{s,K_3(v_2=v_4)})_m,q) + P((G'_{s,K_2(v_1=v_3,v_2=v_4)})_m,q)
\label{addconend}
\eeqs
Let us label the five resultant complete graphs on the right-hand end via the
label $i$, with chromatic polynomials denoted by $P((G'_{s,i})_0,q)$.
  Next, working one's way leftward from the right-hand end,
apply the addition-contraction theorem in a similar manner to
the next set of transverse vertices, and label the five resultant complete
graphs with the label $j$. Define a (square) matrix $M$ (in this case
$5 \times 5$) with elements consisting of the chromatic polynomials
of this set of graphs with complete subgraphs $i$ on the right and $j$ on
the left. This procedure transforms the initial strip into a sum of factorized
strips with parts that overlap in complete graphs. Therefore the intersection
theorem yields, for $m \ge 1$, 
\beq
P((G'_{s,j})_m,q)=\sum_i\frac{M_{ji}P((G'_{s,i})_{m-1},q)}
{P((G'_{s,i})_0,q)}=\sum_i (MD)_{ji}P((G'_{s,i})_{m-1},q),
\label{inteq}
\eeq
where $D$ is a ($5 \times 5$) diagonal matrix with elements
$D_{i,i}=1/P((G'_{s,i})_0,q)$.
Because of (\ref{addconend}), the generating function of $G'_s$ can be
written as
\beq
\Gamma(G'_s,q,x) = \sum_j \Gamma(G'_{s,j},q,x)
\label{gammags}
\eeq
Writing the generating function of $G'_{s,j}$ in the more convenient form
\beq
\Gamma(G'_{s,j},q,x) = \sum_{m=1}^{\infty} P((G'_{s,j})_m,q)x^{m-1}
\label{gammagsj}
\eeq
and using (\ref{inteq}) and (\ref{gammags}), we obtain
\beq
\Gamma(G'_s,q,x)={\bf v_a} (1-x MD)^{-1} M {\bf v_b}={\bf v_a}\sum_{m=0}^
{\infty} x^m (MD)^m M {\bf v_b} ,
\label{gammaaddcon}
\eeq
where in this case, for the strip $G_s'$, ${\bf v_a}=(1,1,1,1,1)$ and
${\bf v_b}={\bf v_a}^t$ denotes the transpose of ${\bf v_a}$. Note that
in general ${\bf v_a}$ and ${\bf v_b}$ can be chosen independently, and
different choices of these vectors correspond to different boundary
conditions on the two ends of the strip.
This approach to the theorem allows one to write (\ref{gamma}) in the
equivalent form
\beq
P((G_s')_m,q)={\bf v_a} (MD)^m M {\bf v_b}
\label{addconpol}
\eeq
The generating function (\ref{gammaaddcon}) is a rational function of
the form (\ref{gammagen}) with the denominator given by
\beq
{\cal D}(G_s',q,x)=\prod_r (1-\lambda_r(q) x),
\label{addconden}
\eeq
where ${\lambda_r(q)}'s$ are eigenvalues of the product of matrices $MD$. The
boundary conditions on the two ends of the strip (defined by ${\bf v_a}$ and
${\bf v_b}$) determine which eigenvalues enter in the product in equation
(\ref{addconden}).  Using these methods, we have also calculated chromatic
polynomials and their asymptotic limits for various strip graphs of the form
$(G_s)_m = J(\prod_{\ell=1}^m H)I$ (with $(G_s)_0 \equiv JI$) where the strip 
has two, in general different, subgraphs on its ends.  We shall present 
these calculations and related discussions of boundary conditions other than 
free ones in a subsequent paper \cite{strip2}.  

\section{Appendix 2} 

   In this Appendix we list the $a_{s,j}$ and $b_{s,k}$ polynomials making up
the generating functions $\Gamma(G_s,q,x)$ for various strip graphs $G_s$.  

\subsection{Strip of the Square Lattice in the Diagonal Direction}

   Here we consider a strip of the square lattice such that the longitudinal
axis of the strip is the diagonal direction of the lattice. In the simplest 
case, where adjacent squares touch each other only at a common vertex, the 
chromatic polynomial factorizes, as $q(q-1)D_4(q)^{m+1}$ for $m$ squares
adjoined to an initial one, $I$.  The first case with a
nonfactorizing chromatic polynomial is for the strip shown in 
Fig.\ \ref{figstrip}(c), 
where the initial subgraph $I$ at the right-hand end is a single diagonally 
oriented square, and the repeating subgraph unit $H$ is a ``wedge'' 
consisting of two squares, one above the other, with a third to the 
immediate left.  We denote the strip graph with $m$ of these repeating units 
$(G_{sq_d(3)})_m$.  
We find a generating function with $j_{max}=1$, $k_{max}=2$, and 
\beq
a_{sq_d(3),0} = q(q-1)D_4(q)
\label{a0sqdwd3}
\eeq
\beq
a_{sq_d(3),1} = -q(q-1)^3(q-2)
\label{a1sqdwd3}
\eeq
\beq
b_{sq_d(3),1} = -q^5+8q^4-28q^3+53q^2-56q+27
\label{b1sqdwd3}
\eeq
\beq
b_{sq_d(3),2} = (q-1)(q-2)D_4(q)^2
\label{b2sqdwd3}
\eeq

   We have also calculated the generating function for a similar lattice of
greater width, shown in Fig.\ \ref{figstrip}(d).  For this case we find 
$j_{max}=2$, $k_{max}=3$, and 
\beq
a_{sq_d(4),0} = q(q-1)D_4(q)^3
\label{a0sqdw4}
\eeq
\beq
a_{sq_d(4),1} = -q(q-1)^2D_4(q)(3q^5-25q^4+86q^3-152q^2+136q-47)
\label{a1sqdw4}
\eeq
\beq
a_{sq_d(4),2} = q(q-1)^3(q-2)^2D_4(q)^3
\label{a2sqdw4}
\eeq
\beq
b_{sq_d(4),1} = -q^6+10q^5-45q^4+116q^3-185q^2+178q-82
\label{b1sqdw4}
\eeq
\beq
b_{sq_d(4),2} = D_4(q)^2(3q^4-22q^3+64q^2-88q+49)
\label{b2sqdw4}
\eeq
\beq
b_{sq_d(4),3} = -(q-2)^2D_4(q)^4
\label{b3sqdw4}
\eeq

\subsection{Strip of the Triangular Lattice of Width $L_{\lowercase{y}}=4$}

   In the main text, we have given our results for the generating function for
a strip of the triangular lattice with width $L_y=3$; here we list the results
for the next wider strip of this type, having a width $L_y=4$.  We calculate
that $j_{max}=3$, $k_{max}=4$, and 
\beq
a_{t(4),0} = q(q-1)(q-2)^6
\label{a0triw3}
\eeq
\beq
a_{t(4),1} = -q(q-1)(q-2)^3(q-3)(3q^4-25q^3+76q^2-98q+43) 
\label{a1triw3}
\eeq
\beq
a_{t(4),2} = q(q-1)(q-2)^6(q-3)^2(3q^2-12q+8)
\label{a2triw3}
\eeq
\beq
a_{t(4),3} = -q(q-1)^3(q-2)^6(q-3)^4
\label{a3triw3}
\eeq
\beq
b_{t(4),1} = -q^4+10q^3-42q^2+88q-76
\label{b1triw3}
\eeq
\beq
b_{t(4),2} = (q-2)(q-3)^2(3q^3-22q^2+60q-60)
\label{b2triw3}
\eeq
\beq
b_{t(4),3} = -(q-2)^2(q-3)^3(3q^3-21q^2+51q-43)
\label{b3triw3}
\eeq
\beq
b_{t(4),4} = (q-2)^6(q-3)^4
\label{b4triw3}
\eeq

\subsection{Strip of the $(3 \cdot 12^2)$ Lattice}

   We consider a strip oriented such that the upper and lower sides of the
12-gons are parallel to the direction of the strip.  The width of the strip is
given by the diameter of one 12-gon and the initial subgraph $I$ on
the right is taken to be a 12-gon and the repeating subgraph unit $H$ is a
12-gon with its two adjacent triangles (which do not touch each other).  
We calculate $j_{max}=1$, $k_{max}=2$ and 
\beq
a_{312(2),0} = q(q-1)D_{12}(q)
\label{a031212}
\eeq
\beq
a_{312(2),1} = -q(q-1)^{10}(q-2)(q^2-5q+5)
\label{a131212}
\eeq
\beq
b_{312(2),1} = -(q-2)(q^9-11q^8+54q^7-156q^6+294q^5-378q^4+336q^3-
204q^2+82q-22)
\label{b131212}
\eeq
\beq
b_{312(2),2} = (q-1)^7(q-2)^3(q^2-5q+5)
\label{b31212}
\eeq

\subsection{Strip of the $(4 \cdot 8^2)$ Lattice}

   The minimal width strip of the $(4 \cdot 8^2)$ lattice is shown in Fig.\ 
\ref{figstrip}(g).  The initial subgraph $I$
is an octagon and the repeating subgraph unit $H$ is an octagon with two 
squares at the same $x$ coordinate (which do not touch each other).  
For this strip we find $j_{max}=1$ and $k_{max}=2$ and calculate 
\beq
a_{488(2),0} = q(q-1)D_8(q)
\label{a0488w1}
\eeq
\beq
a_{488(2),1} = -q(q-1)^6(q-2)(3q^2-9q+7)
\label{a1488w1}
\eeq
\beq
b_{488(2),1} = -q^8+11q^7-55q^6+163q^5-314q^4+406q^3-354q^2+203q-63
\label{b1488w1}
\eeq
\beq
b_{488(2),2} = (q-1)^3(q-2)(3q^2-9q+7)D_4(q)^2
\label{b2488w1}
\eeq

   We have also calculated the generating function for the next wider strip of
the $(4 \cdot 8^2)$ lattice, shown in Fig.\ \ref{figstrip}(h).  
The longitudinal
direction of this strip is rotated $45^\circ$ relative to that of the one
above.  We find $j_{max}=2$, $k_{max}=3$, and calculate 
\beqs
a_{488(3),0} & = & q(q-1)(q^{14}-18q^{13}+153q^{12}-814q^{11}+3030q^{10}
-8358q^9+17655q^8-29104q^7 \nonumber \\ 
  &  &  +37796q^6-38720q^5+31057q^4-19118q^3+8666q^2-2649q+427) 
\label{a0488w2}
\eeqs
\beqs
a_{488(3),1} & = & -q(q-1)^4(q^{15}-23q^{14}+251q^{13}-1719q^{12}+8241q^{11}
-29235q^{10} \nonumber \\ 
&  & +79172q^9-166558q^8+274466q^7-354610q^6+356940q^5 \nonumber \\ 
&  & -275630q^4+158575q^3-64360q^2+16492q-1999)
\label{a1488w2}
\eeqs
\beq
a_{488(3),2} = q(q-1)^{11}(q-2)^2(q-3)^2 D_4(q)^2
\label{a2488w2}
\eeq
\beqs
b_{488(3),1} & = & -q^{12}+16q^{11}-120q^{10}+558q^9-1794q^8+4212q^7
-7437q^6 \nonumber \\
  &  & +10018q^5-10324q^4+8064q^3-4648q^2+1854q-414
\label{b1488w2}
\eeqs
\beqs
b_{488(3),2} & = & (q-1)^4(q^{12}-20q^{11}+188q^{10}-1094q^9+4375q^8
-12640q^7+27033q^6 \nonumber \\
 & & -43164q^5+51235q^4-44380q^3+26931q^2-10462q+2017) 
\label{b2488w2}
\eeqs
\beq
b_{488(3),3} = -(q-1)^8(q-2)^2(q-3)^2 D_4(q)^2
\label{b3488w2}
\eeq

\section{Appendix 3} 

   In this Appendix we include a number of additional figures depicting exact
results on the analytic structure for $W(\{G_s\},q)$ functions for
heteropolygonal Archimedean lattices $G_s$, together with calculations of 
chromatic zeros for long finite strips of each type. 

\eject

\vfill
\eject

\begin{thebibliography}{99}

\bibitem{ice}{W. F. Giauque and J. W. Stout, J. Am. Chem. Soc. {\bf 58}
(1936) 1144).}

\bibitem{lp}{L. Pauling, {\it The Nature of the Chemical Bond}
(Cornell Univ. Press, Ithaca, 1960), p. 466.}

\bibitem{liebwu}{E. H. Lieb and F. Y. Wu, in C. Domb and M. S. Green,
eds., {\it Phase Transitions and Critical Phenomena} (Academic Press,
New York, 1972) v. 1, p. 331.}

\bibitem{ps}{N. G. Parsonage and L. A. K. Staveley, {\it Disorder in Crystals}
(Oxford, 1978).}

\bibitem{atkins}{P. W. Atkins, {\it Physical Chemistry} (Freeman, San
Francisco, 1994), p. 710.}

\bibitem{wannier}{G. H. Wannier, Phys. Rev. {\bf 79} (1950) 357.}

\bibitem{kn}{K. Kano and S. Naya, Prog. Theor. Phys. {\bf 10} (1953) 158.}

\bibitem{scgo}{A. P. Ramirez, G. P. Espinosa, and A. S. Cooper,
Phys. Rev. Lett. {\bf 64} (1990) 2070; C. Broholm, G. Aeppli, G. P. Espinosa,
and A. S. Cooper, {\it ibid.} {\bf 65} (1990) 3173.}

\bibitem{dj}{A. S. Wills, A. Harrison, S. A. M. Mentink, T. E. Mason, and Z.
Tun, ``Magnetic Correlations in Deuteronium Jarosite, a Model $S=5/2$ Kagom\'e
Antiferromagnet'', cond-mat/9607106.}

\bibitem{hafkag}{C. Zeng and V. Elser, Phys. Rev. {\bf B42} (1990) 8436;
R. R. P. Singh and D. Huse, Phys. Rev. Lett. {\bf 68} (1992) 1766;
N. Elstner, R. R. P. Singh, and A. P. Young, {\it ibid} {\bf 71} (1993) 
1629; N. Elstner and A. P. Young, Phys. Rev. {\bf B50} (1994) 6871.}

\bibitem{potts}{R. B. Potts, Proc. Camb. Phil. Soc. {\bf 48} (1952) 106.}

\bibitem{wurev}{F. Y. Wu, Rev. Mod. Phys. {\bf 54} (1982) 235.}

\bibitem{birk}{G. D. Birkhoff, Ann. of Math. {\bf 14} (1912) 42.}

\bibitem{whit}{H. Whitney, Ann. of Math. {\bf 33} (1932) 688;
Bull. Am. Math. Soc. {\bf 38} (1932) 572.}

\bibitem{bl}{G. D. Birkhoff and D. C. Lewis, Trans. Am. Math. Soc.
{\bf 60} (1946) 355.}

\bibitem{rtrev}{R. C. Read, J. Combin. Theory {\bf 4} (1968) 52;
R. C. Read and W. T. Tutte, ``Chromatic Polynomials'',
in {\it Selected Topics in Graph Theory, 3}, eds. L. W. Beineke and
R. J. Wilson (Academic Press, New York, 1988; R. Nelson and R. J. Wilson, 
eds., {\it Graph Colourings} (Longman, Essex, 1990).}

\bibitem{tutterev}{W. T. Tutte {\it Graph Theory}, vol. 21 of
{\it Encyclopedia of Mathematics and its Applications}, ed. Rota,
G. C. (Addison-Wesley, New York, 1984).}

\bibitem{biggsbook}{F. Harary, {\it Graph Theory} (Addison-Wesley, Reading,
1969); N. L. Biggs, { \it Algebraic Graph Theory}, (Cambridge, 
U.K., Cambridge Univ. Press, 1st ed. 1974, 2nd ed. 1993).}

\bibitem{lieb}{E. H. Lieb, Phys. Rev. {\bf 162} (1967) 162.}

\bibitem{baxter}{R. J. Baxter, J. Phys. A {\bf 20} (1987) 5241.  See also 
R. J. Baxter, J. Phys. A {\bf 19} (1986) 2821.}

\bibitem{w}{R. Shrock and S.-H. Tsai, Phys. Rev. {\bf E55} (1997) 5165.} 

\bibitem{w3}{R. Shrock and S.-H. Tsai, J. Phys. A {\bf 30} (1997) 495;
Phys. Rev. {\bf E55} (1997) 6791; Phys. Rev. {\bf E56} (1997) 2733.} 

\bibitem{wa}{R. Shrock and S.-H. Tsai, Phys. Rev. {\bf E56} (1997) 1342; 
Phys. Rev. {\bf E56} (1997) 3935.} 

\bibitem{wn}{R. Shrock and S.-H. Tsai, Phys. Rev. {\bf E56} (1997) 4111.}

\bibitem{sokal}{J. Salas and A. Sokal, J. Stat. Phys. {\bf 86} (1997) 551.}

\bibitem{bds}{N. L. Biggs, R. M. Damerell, and D. A. Sands, J. Combin. Theory B
{\bf 12} (1972) 123; N. L. Biggs and G. H. J. Meredith, J. Combin. Theory B
{\bf 20} (1976) 5. The region diagram for the infinite-$n$ limit of ladder 
graphs, shown as Fig. 1 of the first paper, was not completely correct; the 
right-most 
complex-conjugate curves of the boundary ${\cal B}$ that were thought to 
terminate in region $R_1$ actually do not, but instead continue all the way 
in to the real axis, meeting at $q=2$ and thus enclosing two further 
regions.  See Ref. \cite{read91} and Fig. 3 of Ref. \cite{w}.}

\bibitem{bkw}{S. Beraha, J. Kahane, and N. J. Weiss, J. Comb. Theory B {\bf
28} (1980) 52.}

\bibitem{read91}{R. C. Read and G. F. Royle, in {\it Graph Theory,
Combinatorics, and Applications} (New York, Wiley, 1991), vol. 2, p. 1009; 
R. C. Read, in the {\it Proceedings of the Fifth Caribbean Conference 
on Combinatorics and Computing}, Barbados, 1988. The region diagram for the 
infinite-$n$ limit of ladder graphs, shown as Fig. 4 and 13 in these 
respective papers, was not completely correct; on the left, the
complex-conjugate curves of the boundary ${\cal B}$ that were thought to 
terminate on the unit circle at $q \simeq 0.191 \pm 0.982i$ (i.e., at 
$q=e^{\pm i \theta_e}$ with $\theta_e = \arccos((3-\sqrt{5})/4)$) actually do
not, but instead continue all the way to the real axis, meeting at the origin, 
$q=0$, and thus enclosing another region, as shown in Fig. 3 of Ref. 
\cite{w}. Professor Read has kindly informed us (private communication) that 
he has calculated recursion relations and chromatic zeros for the $L_y=3$ 
square strip in unpublished work.} 

\bibitem{strip2}{M. Ro\v{c}ek, R. Shrock, and S.-H. Tsai, to appear.} 

\bibitem{series}{J. F. Nagle, J. Combin. Theory {\bf 10} (1971) 42; 
G. A. Baker, Jr., J. Combin. Theory {\bf 10} (1971) 217; 
D. Kim and I. G. Enting, J. Combin. Theory, B {\bf 26} (1979) 327.}

\bibitem{gs}{B. Gr\"{u}nbaum and G. Shephard, {\it Tilings and 
Patterns} (Freeman, New York, 1987).}

\bibitem{cmo}{V. Matveev and R. Shrock, J. Phys. A {\bf 28} (1995) 5235.}

\bibitem{z6}{V. Matveev and R. Shrock, Phys. Lett. {\bf A221} (1996) 343.}

\bibitem{alg}{S. Lefschetz, {\it Algebraic Geometry} (Princeton Univ. Press, 
Princeton, 1953); R. Hartshorne, {\it Algebraic Geometry} (Springer, New 
York, 1977).}

\bibitem{is1d}{V. Matveev and R. Shrock, Phys. Lett. {\bf A204} (1995) 353.}

\bibitem{1dnnn}{R. Shrock and S.-H. Tsai, Phys. Rev. {\bf E55} (1997) 5184.}

\bibitem{martin}{P. P. Martin, J. Phys. A {\bf 19} (1986) 3267; 
J. Phys. A {\bf 20} (1986) L601.}

\bibitem{farrell}{E. J. Farrell, Discrete Math. {\bf 29} (1980) 161.} 

\end{thebibliography}
\end{document}